\documentclass[manuscript,screen,review=false]{acmart}
\renewcommand\footnotetextcopyrightpermission[1]{} 
\usepackage{enumitem}
\usepackage{multirow}

\AtBeginDocument{%
  \providecommand\BibTeX{{%
    \normalfont B\kern-0.5em{\scshape i\kern-0.25em b}\kern-0.8em\TeX}}}

\setcopyright{none}
\acmDOI{}

\newcommand{\CorrectSign}{\includegraphics[scale=0.15]{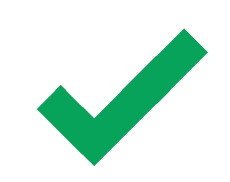}}
\newcommand{\WrongSign}{\includegraphics[scale=0.15]{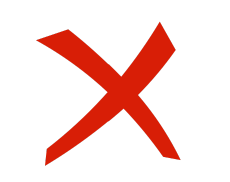}}
\newcommand{\SkipSign}{\includegraphics[scale=0.15]{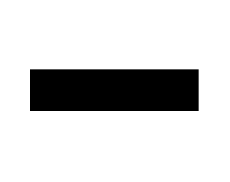}}

\begin{document}

\title{A Question-centric Multi-experts Contrastive Learning Framework for Improving the Accuracy and Interpretability of Deep Sequential Knowledge Tracing Models}

\author{Hengyuan Zhang}
\email{zhang-hy22@mails.tsinghua.edu.cn}
\affiliation{%
  \institution{Shenzhen International Graduate School, Tsinghua University}
  \city{Shenzhen}
  \country{China}}

\author{Zitao Liu}
\authornote{Corresponding author}
\email{liuzitao@jnu.edu.cn}
\affiliation{%
  \institution{Guangdong Institute of Smart Education, Jinan University}
  \city{Guangzhou}
  \country{China}
}
\author{Chenming Shang}
\email{scm22@mails.tsinghua.edu.cn}
\affiliation{%
  \institution{Shenzhen International Graduate School, Tsinghua University}
  \city{Shenzhen}
\country{China}}
\author{Dawei Li}
\email{dal034@ucsd.edu}
\affiliation{%
  \institution{Halicioglu Data Science Institute, University of California, San Diego}
  \city{San Diego}
\country{USA}}

\author{Yong Jiang}
\affiliation{%
  \institution{Shenzhen International Graduate School, Tsinghua University}
  \city{Shenzhen}
  \country{China}
  }








\begin{abstract}
Knowledge tracing (KT) plays a crucial role in predicting students' future performance by analyzing their historical learning processes. 
Deep neural networks (DNNs) have shown great potential in solving the KT problem. 
However, there still exist some important challenges when applying deep learning techniques to model the KT process. The first challenge lies in modeling the individual question information. This is crucial because students' knowledge acquisition on questions that share the same set of knowledge components (KCs) may vary significantly. However, due to the large question bank, the average number of interactions per question may not be sufficient. This limitation can potentially result in overfitting of the question embedding and inaccurate question knowledge acquisition state that relies on its corresponding question representation.
Furthermore, there is a considerable portion of questions receiving relatively less interaction from students in comparison to the majority of questions.
This can further increase the risk of overfitting and lower the accuracy of the obtained question knowledge acquisition state.
The second challenge lies in interpreting the prediction results from existing deep learning-based KT models. 
In real-world applications, while it may not be necessary to have complete transparency and interpretability of the model parameters, it is crucial to present the model's prediction results in a manner that teachers find interpretable.
This makes teachers accept the rationale behind the prediction results and utilize them to design teaching activities and tailored learning strategies for students.
However, the inherent black-box nature of deep learning techniques often poses a hurdle for teachers to fully embrace the model's prediction results.
To address these challenges, we propose a Question-centric Multi-experts Contrastive Learning framework for KT called Q-MCKT. 
This framework explicitly models students' knowledge acquisition state at both the question and concept levels. 
It leverages the mixture of experts technique to capture a more robust and accurate knowledge acquisition state in both question and concept levels for prediction.
Additionally, a fine-grained question-centric contrastive learning task is introduced to enhance the representations of less interactive questions and improve the accuracy of their corresponding question knowledge acquisition states.
Moreover, Q-MCKT utilizes an item response theory-based prediction layer to generate interpretable prediction results based on the knowledge acquisition states obtained from the question and concept knowledge acquisition modules.
We evaluate the proposed Q-MCKT framework on four public real-world educational datasets. 
The experimental results demonstrate that our approach outperforms a wide range of deep learning-based KT models in terms of prediction accuracy while maintaining better model interpretability. 
To ensure reproducibility, we have provided all the datasets and code on our website at \url{https://github.com/rattlesnakey/Q-MCKT}.

\end{abstract}


\nocite{kong2022multitasking,li2023multi,zhang2024improving,shang2024understanding}


\maketitle

\section{Introduction}
\label{sec:intro}
In recent years, the utilization of online educational platforms, such as massive open online courses (MOOCs) and Coursera, has significantly increased, particularly due to the impact of the COVID-19 pandemic and subsequent school closures. 
The surge in online learning has resulted in a wealth of student learning data. 
Furthermore, recent advancements in data analytics have greatly contributed to significant progress in the collection and analysis of extensive learner data.
The availability of vast amounts of learner data, coupled with sophisticated data analytics, has opened up new possibilities for leveraging AI in education settings.
Since every student is unique and has distinct learning needs, personalized learning has emerged as a crucial aspect of education.
As a result, the development of personalized online education platforms, such as intelligent tutoring systems (ITS)~\cite{woolf2010building}, has become a prominent focus within the field of AI in education.
The ability to assess each student's current knowledge level and automatically provide personalized feedback~\cite{piech2015learning} and recommendations~\cite{lan2016contextual} to the students through the large-scale learning data obtained from the online learning environment is the key to the success of ITS.

\begin{figure}[!tbph]
\centering
\includegraphics[width=\columnwidth]{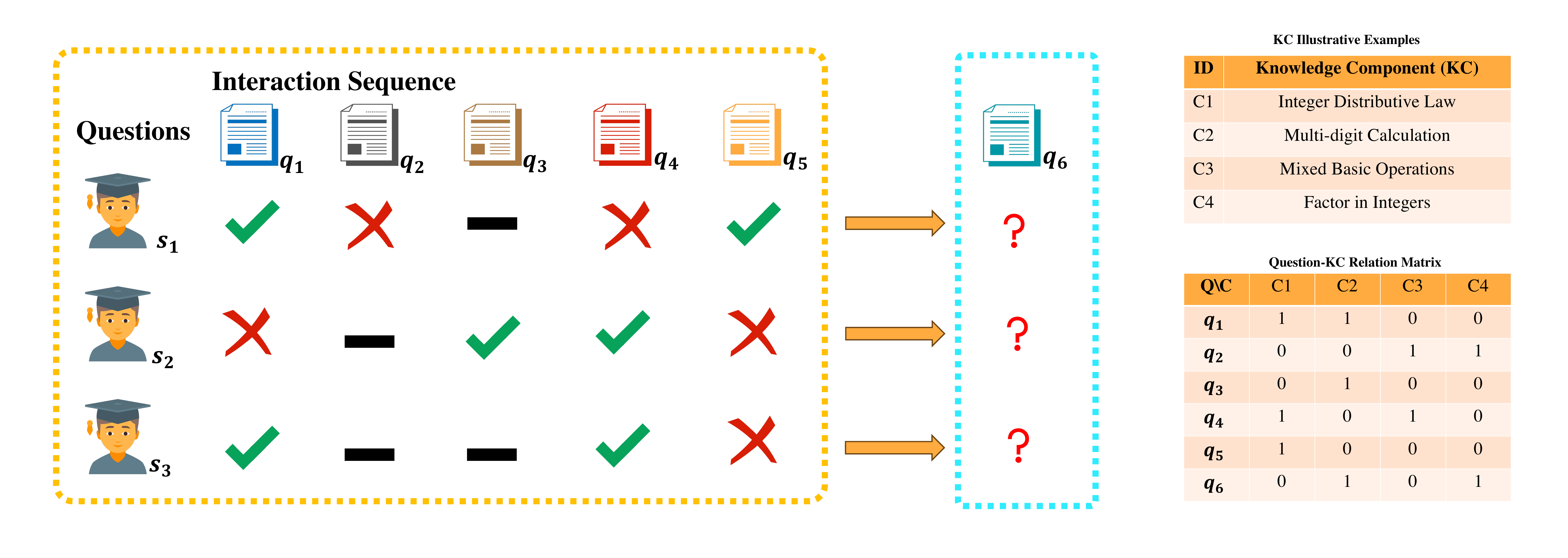}
\caption{A graphical illustration of the KT problem. ``\CorrectSign'' and ``\WrongSign'' denote the question is answered correctly and incorrectly and ``\SkipSign'' denotes the student doesn't get a chance to answer the question.}
\label{fig:kt_illustration}
\end{figure}

Knowledge tracing is an important task in the field of ITS~\cite{cl4kt,qdkt,qikt}. 
It aims to predict the outcomes of students over questions by modeling their mastery of knowledge, i.e., knowledge states, as they interact with learning platforms.
Take Figure~\ref{fig:kt_illustration} as an example, 
student $s_1$ has sequentially answered five questions ($q_1$\textasciitilde$q_5$) where $q_1$ and $q_5$ are answered correctly while $q_2$ and $q_4$ are answered incorrectly.
This indicates that $s_1$ may have a good mastery of the ``Integer Distributive Law'' and ``Multi-digit Calculation'' KCs\footnote{A KC is a generalization of everyday terms like concept, skill, or principle. In the KT task, questions are often associated with one or more KCs, which represent the underlying knowledge that a student needs to grasp in order to answer the question correctly. In this paper, we directly regard ``concept'' as KC.\label{KC}}, but is not familiar with the ``Mixed Basic Operations'' and ``Factor in Integers''. The KCs associated with each question can be found in the \emph{Question-KC Relation Matrix} in Figure~\ref{fig:kt_illustration}. 
Based on the current mastery of knowledge, the KT task can predict the performance of student in the following sixth question $q_6$.
Once the knowledge mastery is acquired through KT, students can find out their weaknesses in time and then carry out targeted exercises.
It can also aid online learning platforms in recommending personalized learning materials to cater to the unique needs of individual students~\cite{wu2020exercise,liu2019ekt}.

The KT related research has been studied since the 1990s when \citet{corbett1994knowledge}, to the best of our knowledge, were the first to estimate students’ current knowledge with regard to each individual KC. 
Since then, many attempts have been made to solve the KT problem, such as probabilistic graphical models~\cite{kaser2017dynamic,hawkins2014learning} and factor analysis-based models~\cite{cen2006learning,lavoue2018adaptive,thai2012factorization}.
While these approaches achieve encouraging progress, they always rely on simplified assumptions such as no forgetting during the learning process~\cite{corbett1994knowledge}, limiting their applications in real scenarios. 
Moreover, these methods are always hand-crafted engineerings. 
However, incorporating all potential factors into the KT model is essentially a non-trivial task.

\begin{figure}[!tbph]
\centering
\includegraphics[width=\columnwidth]{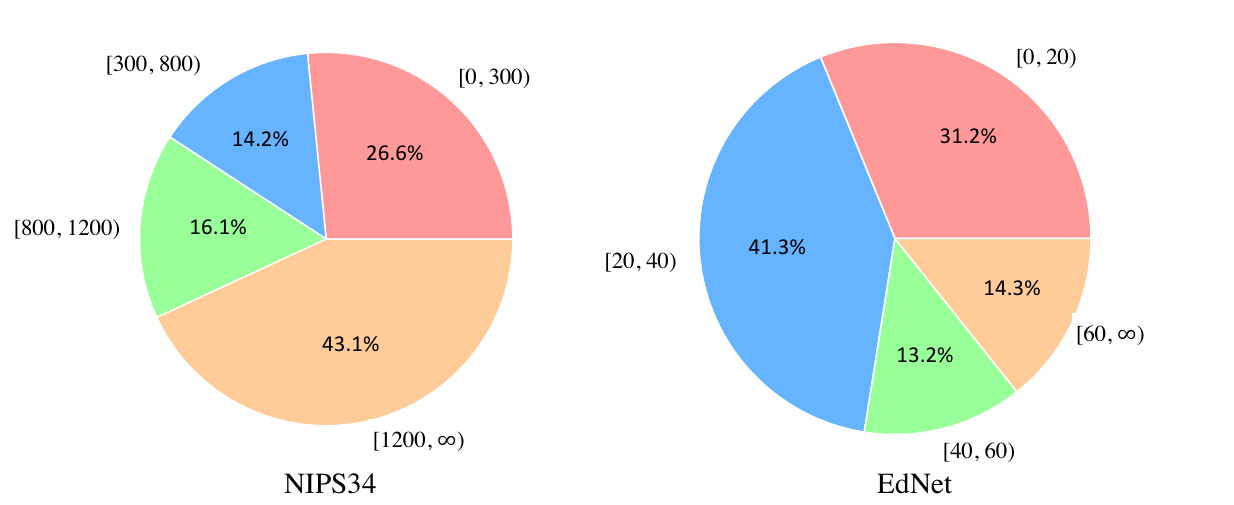}
\caption{The distribution of question interaction frequency in two real-world datasets. }
\label{fig:question_distribution}
\end{figure}

Recently, with the rapid development of deep neural networks, many DLKT models are developed~\cite{qikt,chen2018prerequisite,Adversarial-KT,long2021tracing,dkt,dkt+,abdelrahman2019knowledge,deep-irt,zhang2017dynamic,choi2020towards,AKT,sakt,pandey2020rkt}.
One of the representative approaches among them is the deep sequential KT modeling, which utilizes auto-regressive architectures, such as LSTM and GRU~\cite{lstm,gru}, to represent student’s knowledge states (e.g., the mastery level of KCs) as the hidden states of recurrent units~\cite{qikt,chen2018prerequisite,Adversarial-KT,long2021tracing,dkt,dkt+}. 
Due to the ability to learn sequential dependencies from student interaction data, deep sequential KT models draw attention from researchers from different communities and achieve great success in improving the KT prediction accuracy~\cite{minn2018deep,nagatani2019augmenting,su2018exercise}.
In spite of the promising results demonstrated by previous methods, some important challenges still exist when applying deep sequential KT models on real-world educational data. 
Firstly, taking the individual information of the question into modeling is challenging for the KT model. 
This challenge is caused by the large question bank, which leads to an average number of interactions per question may not be sufficient.
Thus, it can potentially result in overfitting of the question embedding and inaccurate question knowledge acquisition state\footnote{Question knowledge acquisition state denotes the student's knowledge acquisition level for each question within the question bank, it is obtained by the corresponding question embedding. More details are illustrated in Section~\ref{sec:MQKA}.\label{q_acq}} that relies on their corresponding question representation.
Additionally, there exists a considerable portion of less interactive questions in the dataset, which have a higher risk of getting overfitted.
For example, as shown in Figure~\ref{fig:question_distribution}, more than 31\% of questions are associated with an interaction frequency of less than 20 in the EdNet dataset, and 25\% of questions have an interaction frequency of less than 300 in the NIPS34 dataset.
Consequently, it is difficult for KT model to accurately model the question knowledge acquisition state, especially when encountering the less interactive question.
Secondly, despite the significant improvements in prediction accuracy achieved by DLKT models compared to traditional cognitive models, there is still a challenge in extracting psychologically interpretable prediction results from them.  
This lack of interpretability often leaves teachers dissatisfied, as they require interpretable prediction results before assigning appropriate learning materials for students.
It should be noted that although the model parameters may not be explainable, the reasoning behind the prediction results must be interpretable for teachers to accept.
\begin{figure}[!tbph]
\centering
\includegraphics[width=\columnwidth]{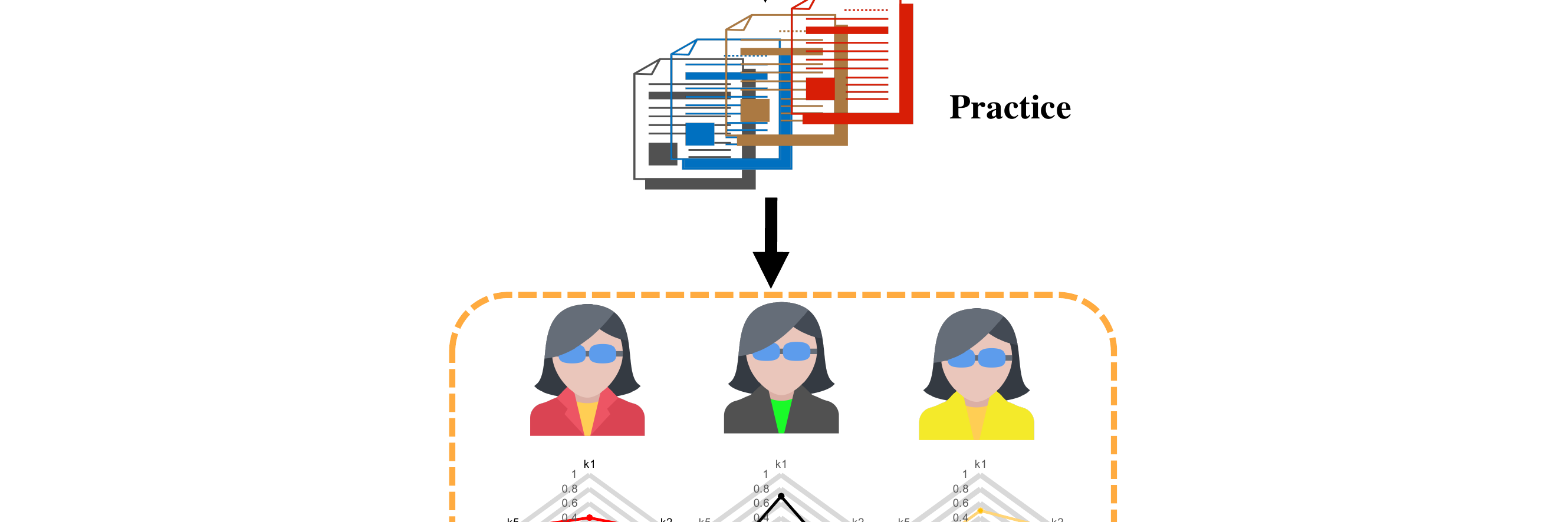}
\caption{A visualization showcasing the overall scores given by multiple teachers.}
\label{fig:MoE}
\end{figure}

In this paper, we address aforementioned challenges by proposing a novel framework called \textbf{Q}uestion-centric \textbf{M}ulti-experts \textbf{C}ontrastive Learning framework for \textbf{KT}, i.e., Q-MCKT. 
Specifically, the question knowledge acquisition module of Q-MCKT estimates the question knowledge acquisition state\textsuperscript{\ref{q_acq}} of students by modeling the variations in students’ question knowledge states after receiving responses to specific questions.
To mitigate the potential inaccuracy of the question knowledge acquisition state caused by the overfitted question embedding, we draw inspiration from the realistic educational scenario where multiple teachers evaluate a student's performance to obtain an overall score (as illustrated in Figure~\ref{fig:MoE}) and incorporate the mixture of experts technique~\cite{MoE} into Q-MCKT to obtain a more robust and accurate question knowledge acquisition state.
For less interactive questions, Q-MCKT further employs an auxiliary fine-grained question-centric contrastive learning procedure to enhance its question representations and thereby improves the accuracy of its corresponding question knowledge acquisition state.
Furthermore, in real educational settings, students often make guesses or slip-ups when interacting with certain questions, which can lead the question knowledge acquisition state overly sensitive to individual interactions.
Therefore, Q-MCKT also incorporates a concept knowledge acquisition module in a similar manner to the question knowledge acquisition module.
This module aims to obtain the concept knowledge acquisition state\footnote{Concept knowledge acquisition state denotes the student's acquisition level for each concept. It is obtained by the concept embedding, more details are illustrated in Section~\ref{sec:MCKA}.}, which serves as the supplemental information for the question knowledge acquisition state.
To enhance the interpretability of prediction results, Q-MCKT directly feeds the knowledge acquisition states from question and concept knowledge acquisition modules into an interpretable prediction layer.
This interpretable prediction layer utilizes the Item Response Theory (IRT)~\cite{irt} from psychometrics, which is widely used in educational scenarios.
We evaluate Q-MCKT on four benchmark datasets by comparing it with 15 previous approaches under a rigorous KT evaluation protocol~\cite{liu2022pykt}.
The experimental results demonstrate that Q-MCKT achieves superior prediction performance and psychologically meaningful interpretability simultaneously. 

The main contributions are summarized as follows:

\begin{itemize}[leftmargin=0.8cm]

\item We simultaneously model the knowledge acquisition at question and concept levels, and we leverage the mixture of experts technique to make its corresponding knowledge acquisition states for the interpretable prediction layer more robust and accurate. Furthermore, we design a fine-grained question-centric contrastive learning approach to enhance the representation of the less interactive question and thereby mitigate its negative impact on the corresponding question knowledge acquisition state.
\item We design a simple yet effective interpretable prediction layer based on the IRT theory and manage to seamlessly combine it with existing deep sequential KT models.
\item We conduct quantitative and qualitative experiments to validate the performance of Q-MCKT on four public datasets with a wide range of baselines. The well-designed experiments illustrate the superiority of our approach in both prediction performance and model interpretability. 

\end{itemize}

\section{Related Work}
\label{sec:related}
\subsection{Knowledge Tracing}
There exist two paradigms of knowledge tracing techniques: conventional methods and deep learning-based methods.
When it comes to conventional methods, one of the most renowned methods is Bayesian knowledge tracing (BKT)~\cite{bkt2}. BKT utilized binary variables to depict the knowledge state of students and leveraged the Hidden Markov Model to capture the sequential progression of knowledge states. 
Subsequent studies have delved into integrating additional factors (e.g. question difficulty~\cite{bkt4}, guess estimation~\cite{bkt5} and student ID~\cite{bkt3}) related to learning scenarios into the modeling process. 
These investigations have revealed that as the number of incorporated modeling factors grows, the performance of the models improves correspondingly. 
As deep learning techniques have achieved great success in various applications, it has also been applied in KT problem to model student’s historical learning interactions and enable a comprehensive understanding of the dependencies within the learning interactions, leading to enhanced predictive capabilities.
The recent deep learning-based methods can be categorized as follows:

\begin{itemize}[leftmargin=*]
    \item \textbf{Deep sequential models}: the chronologically ordered interaction sequence is captured by deep sequential models such as LSTM and GRU \cite{chen2018prerequisite,Adversarial-KT,lee2019knowledge,liu2019ekt,minn2018deep,nagatani2019augmenting,dkt,su2018exercise,dkt+}.
    For example, \citet{dkt} proposed deep knowledge tracing (DKT) model, which employed the hidden state of LSTM to represent student's knowledge state and then fed the hidden state into a binary classifier to predict the students' response performances.
    \citet{dkt+} utilized prediction-consistent regularization technique to address the problems of input reconstruction failure and prediction inconsistency in DKT~\cite{dkt}.
    Motivated by the learning curve theory~\cite{learning_curve},~\citet{nagatani2019augmenting} taked student's forgetting behavior into consideration to enhance DKT~\cite{dkt}. 
    \citet{lee2019knowledge} used student knowledge state encoder and skill encoder to predict the student response performance via the dot product.
    
    \item \textbf{Memory augmented models}: the latent relationships among KCs are effectively captured through the utilization of an external memory structure, especially key-value memory, which is updated iteratively \cite{abdelrahman2019knowledge,shen2021learning,zhang2017dynamic}. 
    For example, \citet{zhang2017dynamic} incorporated a well-designed static key matrix to store the interconnections between various KCs, while simultaneously employing a dynamic value matrix to iteratively update the students' knowledge state.

    \item \textbf{Adversarial based models}: the adversarial training techniques such as adversarial perturbations are applied into the original student interaction sequence to reduce the risk of DLKT overfitting and limited generalization problem. 
    For example, \citet{Adversarial-KT} enhanced the generalization ability of the DLKT model by introducing adversarial perturbations into the student interaction sequence at embedding level. Due to its carefully crafted perturbations into the sequence, the augmenting model can generalize well to diverse student interactions.
    
    \item \textbf{Graph based models}: the response interactions between students and questions and the knowledge associations between questions and KCs form a tripartite graph and graph based techniques are applied to aggregate such relations \cite{nakagawa2019graph,tong2020structure,yang2020gikt}.
    For example, \citet{nakagawa2019graph} utilized graph neural network to aggregate the node features related to the answered concept and subsequently updated the student's knowledge states.

    \item \textbf{Attention based models}: dependence between interactions is captured by the attention mechanism and its variants \cite{AKT,pandey2020rkt,pu2020deep,zhang2021multi}. 
    For example, \citet{AKT} leveraged an attention mechanism to characterize the time distance between questions and the past interaction of students.
    \citet{sakt} utilized a self-attention mechanism to capture relations between exercises and the student responses.
    \citet{pandey2020rkt} introduced a relation-aware self-attention layer that taked the individual question textual content and student past performance to rich the contextual information of student learning process.
    \citet{choi2020towards} used the Transformer-based encoder-decoder architecture to capture students' exercise and response sequences.
    
\end{itemize}

Please note that the above categorizations are not exclusive and related techniques can be combined. 
For example, \citet{abdelrahman2019knowledge} proposed a sequential key-value memory network to unify the strengths of recurrent modeling capacity and memory capacity.

\subsection{Interpreting Deep Learning Based Knowledge Tracing Models}
In recent years, an increasing number of interpretable techniques have been integrated into DLKT models, enhancing both student modeling and prediction tasks. 
These techniques can be categorized into three distinct categories, as follows:

\begin{itemize}[leftmargin=*]
\item \textbf{C1: Post-hoc local explanation}. Post-hoc local explanation techniques are employed in the KT task to scrutinize each individual prediction and understand the underlying reasoning behind the decisions made by DLKT models \cite{lu2020towards,lu2022interpreting}. For example, \citet{lu2022interpreting} utilized a layer-wise relevance propagation method, which leveraged the back propagating relevance scores from the model's output layer to interpret the deep sequential KT model.
\item \textbf{C2: Global interpretability with explainable structures}. Aggregate an interpretable cognitive module into existing DLKT architectures to provide a transparent and interpretable representation of the student's knowledge state \cite{wang2020neural,zhao2020interpretable,pu2022eakt}. For example, \citet{wang2020neural} incorporated an intermediate interaction layer based on multidimensional IRT and model both student factors and exercise factors, allowing for a more fine-grained understanding of the knowledge state modeling process. \citet{pu2022eakt} introduced an automatic temporal cognitive method to enhance the modeling of changes in students' knowledge states over time.  
\item \textbf{C3: Global interpretability with explainable parameters}. Get the explainable parameters obtained from outputs of the DLKT models and directly feed it into cognitively interpretable models to estimate the probability that a student will answer a question correctly \cite{converse2021incorporating,yeung2019deep,qikt}. For example, \citet{qikt} cast the parameters from problem solving and question-agnostic knowledge mastery modules into the IRT-like form. \citet{yeung2019deep} modeled the levels of student abilities and KC difficulties with a dynamic key-value memory network for KT task and fed the learned results to an IRT layer for final prediction. 
\end{itemize}

Our Q-MCKT belongs to the C3 category since we utilize the IRT function for interpretable prediction. 
Unlike existing approaches \cite{yeung2019deep,converse2021incorporating} that solely focus on optimizing the model performance based on the outcomes of the interpretable prediction layer, our Q-MCKT take the explainable parameter learning into the final model optimization objective, which improves the model interpretability and preserves the prediction performance as well.  
Compared to methods developed using memory networks (e.g. Deep-IRT \cite{yeung2019deep}), our approach is based on deep sequential architectures, which offer greater applicability and improved prediction accuracy. 

\subsection{Constrastive Learning}
Contrastive learning incorporates two major procedures: the construction of positive and negative pairs and the calculation of the contrastive loss.
The contrastive loss function measures the similarity between the representations of positive pairs and encourages their embeddings to be close in the feature space. 
Simultaneously, it enforces the representations of negative pairs to be distant from each other. 
This process assists model in learning more discriminative and informative representations.
As contrastive learning has achieved great success in the realm of representation learning, its application has extended to various downstream scenarios.
For example, \citet{simcse} used a dropout trick to derive positive samples in the embedding level, and then applied both supervised and self-supervised methods to acquire better sentence embedding.
\citet{radford2021learning,li2020oscar} used contrastive learning to pre-train a vision language model to align the message between images and their corresponding text.
~\citet{zhang2022fine} regarded word and definition as a semantic equivalence pair to do contrastive learning to enhance the word and definition representations.
\citet{zhang2023assisting} implemented contrastive learning as an auxiliary task to obtain a more oriented and pure task prompt representation.
\citet{gunel2020supervised} designed an auxiliary contrastive loss to enhance the generalization performance of rare examples and improve the robustness of the model.
In the KT task, \citet{cl4kt} proposed CL4KT framework. They did automatic data augmentation (e.g. question masking, interaction cropping, and interaction permutation) on the original student learning sequence to obtain the positive and negative pairs of learning sequence.
Subsequently, it pooled these sequence representation into an overall dense embedding to conduct contrastive learning. 
However, the positive and negative pairs are built at the learning sequence level rather than the direct question level, and the pairs generated by the automatic data augmentation method lack the necessary semantic alignment information, which can lead to inaccuracies and inconsistencies when doing contrastive learning.
Furthermore, the pooling operation used to acquire representations for computing batch contrastive loss results in the loss of sequence information, thereby adversely affecting question-level representation learning during contrastive learning.
In this work, we carefully design question-centric positive and negative pairs that encompass semantic alignment information at a fine-grained level.
Additionally, our proposed method directly learns question-level representations without relying on pooling operations, making the learning procedure more effective.
This thoughtful design aims to enhance the representation of the less interactive question, thereby leading to a improvement in accurately getting its question knowledge acquisition states for interpretable prediction layer.

\subsection{Mixture of Experts}
The mixture of experts technique comprises two key components: multiple expert networks and a gating network.
The expert networks focus on capturing patterns or information from different perspectives, while the gating network acts as a decision-maker, determining the importance or contribution of each expert's output.
By combining the expertise of different expert networks, the mixture of experts technique can leverage the strengths of each expert to enhance the overall performance.
Recently, the mixture of experts technique has gained significant popularity in various domains, such as computer vision, natural language processing, and recommender systems \cite{lepikhin2020gshard,fedus2022switch,lewis2021base,du2022glam,artetxe2021efficient,ryabinin2020towards,riquelme2021scaling,liang2021credit,alshaikh2020mixture}.
Since the mixture-of-experts is very conducive to parallelization, some works have taken this technique to build the large-scale language and vision models \cite{lepikhin2020gshard,fedus2022switch,lewis2021base,du2022glam,artetxe2021efficient,ryabinin2020towards,riquelme2021scaling}.
By employing a gating network to select the most suitable expert during the forward pass, these models can scale up their parameters without incurring additional computational overhead.
Moreover, the mixture of experts has been leveraged to obtain more comprehensive and robust representations or prediction results \cite{liang2021credit,alshaikh2020mixture,MoE}.
In our work, we adopt the mixture of experts technique to enhance the robustness and accuracy of the knowledge acquisition states for the interpretable prediction layer.

\section{Problem Statement}
\label{sec:ps}
Our objective is to give an arbitrary question $q_*$ to predict the probability of whether a student will answer $q_*$ correctly or not based on the student's historical interaction data. More specifically, for each student $\mathbf{S}$, we assume that we have observed a chronologically ordered collection of $T$ past interactions i.e., $\mathbf{S} = \{\mathbf{s}_t\}_{t=1}^T$. Each interaction is represented as a 4-tuple $\mathbf{s}_t$, i.e., $\mathbf{s}_t = <q_t, \{k\}_t, r_t, t_t>$, where $q_t, \{k\}_t, r_t, t_t$ represent the specific question, the associated KC set, the binary valued student response, and student's response time step respectively. The response is a binary valued indicator variable where 1 represents the student correctly answered the question, and 0 otherwise. We would like to estimate the probability $\hat{r}_{*}$ of the student's performance on the arbitrary question $q_*$.

\section{The Framework}
\label{sec:method}
\begin{figure}[!tbph]
\centering
\includegraphics[width=\columnwidth]{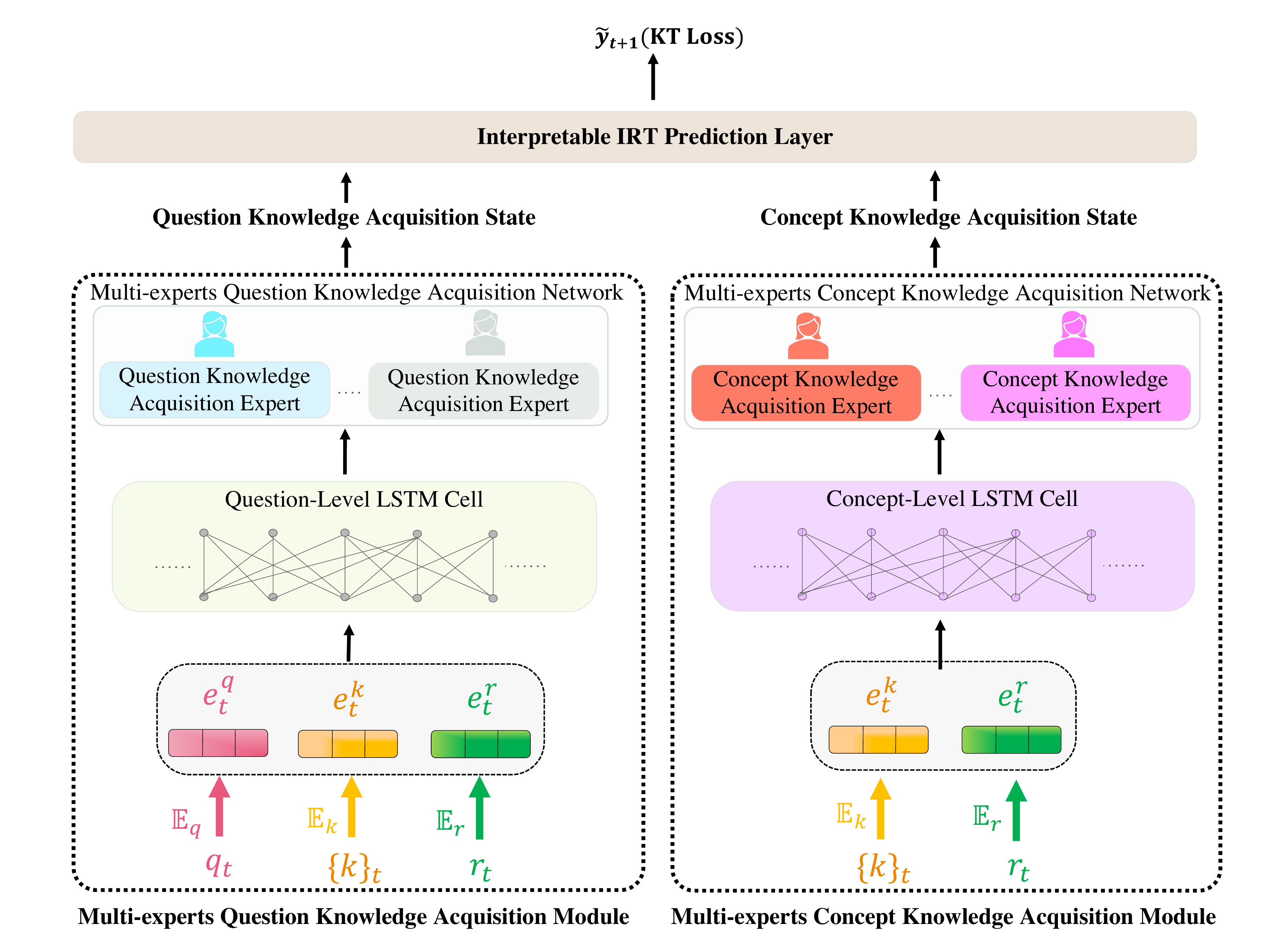}
\caption{The graphical illustrations of prediction modeling in our Q-MCKT model.}
\label{fig:overview}
\end{figure}

In this section, we introduce the components in our proposed \textit{Q-MCKT} framework in detail: 
(1) the Interaction Encoder that encodes both question and concept level information (\ref{sec:interaction_projection}); 
(2) the Multi-experts Question Knowledge Acquisition (MQKA) module that examines the acquisition state of students' knowledge in questions (\ref{sec:MQKA}); 
(3) the Multi-experts Concept Knowledge Acquisition (MCKA) module that examines the acquisition state of students' knowledge in concepts (\ref{sec:MCKA}); 
(4) the Fine-grained Question-centric Contrastive Learning that enhances the representation of the less interactive question to further improve the performance of getting a more accurate question knowledge acquisition state (\ref{sec:cl});
and (5) the Interpretable Prediction Layer that leverages the psychological theory of IRT to generate more interpretable results (\ref{sec:prediction_layer}); 
\subsection{Interaction Encoder}
\label{sec:interaction_projection}
Interaction projection methods such as qDKT~\cite{qdkt}, which neglect the relationships between questions and KCs and rely solely on the question sequence to track students' knowledge states, may not provide an adequate representation of interaction.

Therefore, we follow \citet{long2021tracing} to encode the interaction in a more granular manner by capturing the relationship between questions and KCs.
In our methodology, we employ separate raw interaction encodings $\mathbf{e}_t$ and $\mathbf{c}_t$ for the MQKA module and the MCKA module respectively. 
Specifically, in the MQKA module, we represent each question level interaction $\mathbf{e}_t$ as a concatenation of the question, the student's response, and the corresponding set of KCs associated with the current question, i.e., 

\begin{align}
	\label{eq:et}
	\mathbf{e}_{t}=\left\{
		\begin{array}{l}
			\mathbf{e}^{q}_{t} \oplus \bar{\mathbf{e}}^{k}_{t} \oplus \mathbf{0}, \quad r_{t}=1 \\
			\mathbf{0} \oplus \mathbf{e}^{q}_{t} \oplus \bar{\mathbf{e}}^{k}_{t}, \quad  r_{t}=0 \\
		\end{array}\right.
\end{align}

\noindent where $\mathbf{e}^{q}_{t}$ is the question embedding, $\mathbf{e}^{q}_{t} \in \mathbb{R}^{d \times 1}$ and $\bar{\mathbf{e}}^{k}_{t}$ is the average embeddings of all the KCs to the question, i.e., 

\begin{equation*}
	\bar{\mathbf{e}}^{k}_{t} = \frac{1}{m} \sum_{m = 1}^m \mathbf{e}^{k_m}_{t} * \mathbb{I}(k_m \in \{k\}_t)
\end{equation*}

\noindent where $\mathbf{e}^{k_m}_{t} \in \mathbb{R}^{d \times 1}$ is the KC embedding. $m$ is the total number of KCs in the question $q_t$. 
$\mathbb{I}(\cdot)$ is the indicator function and $\oplus$ is the concatenate operation. The response in each interaction is encoded as a $2d \times 1$ all-zero vector, $\mathbf{0}$. We use concatenation directions (left or right) to indicate different responses, i.e., correct or wrong.

In the MCKA module, as it only estimates the mastery of concept knowledge acquisition, we ignore the question relevant information when encoding the concept level interaction $\mathbf{c}_{t}$: 

\begin{align}
	\label{eq:ct}
	\mathbf{c}_{t}=\left\{
		\begin{array}{l}
			\bar{\mathbf{e}}^{k}_{t} \oplus \mathbf{0}, \quad r_{t}=1 \\
			\mathbf{0} \oplus \bar{\mathbf{e}}^{k}_{t}, \quad  r_{t}=0 \\
		\end{array}\right.
\end{align}

\subsection{Multi-experts Question Knowledge Acquisition Module}
\label{sec:MQKA}
Students absorb question knowledge as they interact with questions and their question knowledge acquisition varies after solving the heterogeneous questions. 
Hence, we propose to estimate students' question knowledge acquisition at timestamp $t$ with the interaction embedding sequence $\{\mathbf{e}_0, \mathbf{e}_1, ...\mathbf{e}_t\}$. 
Similar to the standard DKT model, we use the LSTM cell to update the student's question knowledge state $\mathbf{a}_t$ after answering each question at timestamp $t$:
\begin{align*}
\mathbf{i}_{t} & = \sigma\left(\mathbf{W}_1 \cdot \mathbf{e}_t +\mathbf{U}_1 \cdot \mathbf{a}_{t-1}+\mathbf{b}_1\right) \\
\mathbf{f}_{t} & = \sigma\left(\mathbf{W}_2 \cdot \mathbf{e}_t +\mathbf{U}_2 \cdot \mathbf{a}_{t-1}+\mathbf{b}_2\right) \\
\mathbf{o}_{t} & = \sigma\left(\mathbf{W}_3 \cdot \mathbf{e}_t +\mathbf{U}_3 \cdot \mathbf{a}_{t-1}+\mathbf{b}_3\right) \\
\mathbf{\tilde{c}}_{t} & = \sigma\left(\mathbf{W}_4 \cdot \mathbf{e}_t +\mathbf{U}_4 \cdot \mathbf{a}_{t-1}+\mathbf{b}_4\right) \\
\mathbf{c}_{t} & = \mathbf{f}_{t} \odot \mathbf{c}_{t-1}+\mathbf{i}_{t} \odot \mathbf{\tilde{c}}_{t} \\
\mathbf{a}_t & = \mathbf{o}_{t} \odot \tanh \left(\mathbf{c}_{t}\right) 
\end{align*}

\noindent where $\mathbf{W}_i$s, $\mathbf{U}_i$s, $\mathbf{b}_i$s are trainable parameters and $\mathbf{W}_i \in \mathbb{R}^{d \times 4d}$, $\mathbf{U}_i \in \mathbb{R}^{d \times d}$, $\mathbf{b}_i \in \mathbb{R}^{d \times 1}$ and $i = 1, 2, 3, 4$. $\sigma$, $\odot$, and $\tanh$ denote the sigmoid, element-wise multiplication and hyperbolic tangent functions.

To mitigate the inaccuracy of the question knowledge acquisition state caused by the potential overfitted question embedding, we propose a multi-experts question knowledge acquisition network.
This network is constructed based on the widely recognized mixture of experts technique~\cite{MoE}. 
By leveraging this technique, we can enhance the robustness and accuracy of the question knowledge acquisition state.

Specifically, it is composed of a multi-experts question knowledge acquisition network and a gating network. 
The former contains several expert modules to capture the student's question knowledge acquisition from different perspectives. 
The latter computes a softmax vector to assign weight to each expert module, indicating their respective importance.
To get each expert's question knowledge acquisition state $\mathbf{\gamma}^{q_{e}}_{t}$, each expert first extracts from $\mathbf{a}_t$ with a fully connected neural layer and then project it into the question space via non-linear transformation, i.e.:
\begin{equation}
     \mathbf{\gamma}^{q_{e}}_{t} = \mathbf{w}^a \odot \mbox{ReLU} \bigl( \mathbf{W}^a_2 \cdot \mbox{ReLU} ( \mathbf{W}^a_1 \cdot \mathbf{a}_t + \mathbf{b}^a_1 ) + \mathbf{b}^a_2 \bigl) 
\end{equation}
\noindent where $\mathbf{\gamma}^{q_{e}}_{t} \in \mathbb{R}^{n \times 1}$ is the question knowledge acquisition state of expert $e$, $n$ is the total number of questions, the p-th element of $\mathbf{\gamma}^{q_{e}}_{t}$ denotes the estimated acquisition of student in question $p$. 
$\mathbf{W}^a_1$, $\mathbf{W}^a_2$, $\mathbf{w}^a$, $\mathbf{b}^a_1$ and $\mathbf{b}^a_2$ are trainable parameters and $\mathbf{W}^a_1 \in \mathbb{R}^{d \times d}$, $\mathbf{W}^a_2 \in \mathbb{R}^{n \times d}$, $\mathbf{w}^a \in \mathbb{R}^{n \times 1}$, $\mathbf{b}^a_1 \in \mathbb{R}^{d \times 1}$, $\mathbf{b}^a_2 \in \mathbb{R}^{n \times 1}$.

Finally, the overall question knowledge acquisition state $\mathbf{\gamma}^q_{t} \in \mathbb{R}^{n \times 1}$ is computed as:
\begin{equation}
    \begin{split}
        \mathbf{\gamma}^q_{t} &= \sum_{e=1}^{E} g^{q_{e}}_{t} * \mathbf{\gamma}^{q_{e}}_{t} \\
        \mathbf{g}^q_{t} &= g^q(\mathbf{a}_t) \\
    \end{split}
\end{equation}
\noindent where $g^q$ is the gating network, $\mathbf{g}^q_t \in \mathbb{R}^{e \times 1}$ is the gating vector, $g^{q_{e}}_{t}$ denotes the e-th element of $\mathbf{g}^q_t$, $E$ is the number of experts.

\subsection{Multi-experts Concept Knowledge Acquisition Module}
\label{sec:MCKA}
However, in real educational settings, students often make guesses or slip-ups when interacting with certain questions. 
This can lead to the MQKA module becoming overly sensitive to individual interactions, resulting in inaccuracies during the inference stage.
To mitigate the issue of inaccuracies caused by guesses or slip-ups, we introduce a supplemental knowledge acquisition, namely concept knowledge acquisition.
Concept knowledge acquisition can be regarded as a more general acquisition of knowledge, because the interactions of student in specific concept is much more than in specific question.

To model concept knowledge acquisition, we propose the multi-experts concept knowledge acquisition module. 
Similar to the question knowledge acquisition modeling in the MQKA module, we utilize an additional LSTM cell to update the student's concept knowledge state $\mathbf{v}_t$ after receiving each response.
The LSTM cell in the MCKA module takes concept-level interaction $\mathbf{c}_t$ instead of $\mathbf{e}_t$:
\begin{align*}
    \mathbf{i}_{t} & = \sigma\left(\mathbf{W}_5 \cdot \mathbf{c}_t +\mathbf{U}_5 \cdot \mathbf{v}_{t-1}+\mathbf{b}_5\right) \\
    \mathbf{f}_{t} & = \sigma\left(\mathbf{W}_6 \cdot \mathbf{c}_t +\mathbf{U}_6 \cdot \mathbf{v}_{t-1}+\mathbf{b}_6\right) \\
    \mathbf{o}_{t} & = \sigma\left(\mathbf{W}_7 \cdot \mathbf{c}_t +\mathbf{U}_7 \cdot \mathbf{v}_{t-1}+\mathbf{b}_7\right) \\
    \mathbf{\tilde{c}}_{t} & = \sigma\left(\mathbf{W}_8 \cdot \mathbf{c}_t +\mathbf{U}_8 \cdot \mathbf{g}_{t-1}+\mathbf{b}_8\right) \\
    \mathbf{c}_{t} & = \mathbf{f}_{t} \odot \mathbf{c}_{t-1}+\mathbf{i}_{t} \odot \mathbf{\tilde{c}}_{t} \\
    \mathbf{v}_t & = \mathbf{o}_{t} \odot \tanh \left(\mathbf{c}_{t}\right) 
\end{align*}
    
\noindent where $\mathbf{W}_5$, $\mathbf{W}_6$, $\mathbf{W}_7$, $\mathbf{W}_8$, $\mathbf{U}_5$, $\mathbf{U}_6$, $\mathbf{U}_7$, $\mathbf{U}_8$, $\mathbf{b}_5$, $\mathbf{b}_6$, $\mathbf{b}_7$, $\mathbf{b}_8$ are trainable parameters and $\mathbf{W}_i \in \mathbb{R}^{d \times 3d}$, $\mathbf{U}_i \in \mathbb{R}^{d \times d}$, $\mathbf{b}_i \in \mathbb{R}^{d \times 1}$ and $i = 5, 6, 7, 8$. 
    
Furthermore, similar to MQKA, we also design a multi-experts concept knowledge acquisition network to capsule the overall concept acquisition knowledge state $\mathbf{\gamma}^c_{t}$.
Specifically, each concept knowledge acquisition expert get its corresponding acquisition state $\mathbf{\gamma}^{c_{e}}_{t}$ from $\mathbf{v}_t$ with a fully connected neural layer and then project it into the concept space via non-linear transformation, i.e.:

\begin{equation}
    \mathbf{\gamma}^{c_{e}}_{t} = \mathbf{w}^v \odot \mbox{ReLU} \bigl( \mathbf{W}^v_2 \cdot \mbox{ReLU} ( \mathbf{W}^v_1 \cdot \mathbf{v}_t + \mathbf{b}^v_1 ) + \mathbf{b}^v_2 \bigl) 
\end{equation}
\noindent where $\mathbf{\gamma}^{c_{e}}_{t} \in \mathbb{R}^{m \times 1}$ is the concept knowledge acquisition state of expert $e$, $m$ is the total number of concepts, the p-th element of $\mathbf{\gamma}^{c_{e}}_{t}$ denotes the estimated acquisition of student in concept $p$.
where $\mathbf{W}^v_1$, $\mathbf{W}^v_2$, $\mathbf{w}^v$, $\mathbf{b}^v_1$ and $\mathbf{b}^v_2$ are trainable parameters and $\mathbf{W}^v_1 \in \mathbb{R}^{d \times d}$, $\mathbf{W}^v_2 \in \mathbb{R}^{m \times d}$, $\mathbf{w}^v \in \mathbb{R}^{m \times 1}$, $\mathbf{b}^v_1 \in \mathbb{R}^{d \times 1}$, $\mathbf{b}^v_2 \in \mathbb{R}^{m \times 1}$.

Finally, the overall concept knowledge acquisition state $\mathbf{\gamma}^c_{t} \in \mathbb{R}^{m \times 1}$ is computed as:
\begin{equation}
   \begin{split}
       \mathbf{\gamma}^c_{t} &= \sum_{e=1}^{E} g^{c_{e}}_{t} * \mathbf{\gamma}^{c_{e}}_{t} \\
       \mathbf{g}^c_{t} &= g^c(\mathbf{v}_t) \\
   \end{split}
\end{equation}
\noindent where $g^c$ is the gating network, $\mathbf{g}^c_t \in \mathbb{R}^{e \times 1}$ is the gating vector, $g^{c_{e}}_{t}$ denotes the e-th element of $\mathbf{g}^c_t$.

\subsection{Fine-grained Question-centric Contrastive Learning}
\label{sec:cl}

\begin{figure}[!tbph]
    \centering
    \includegraphics[width=\columnwidth]{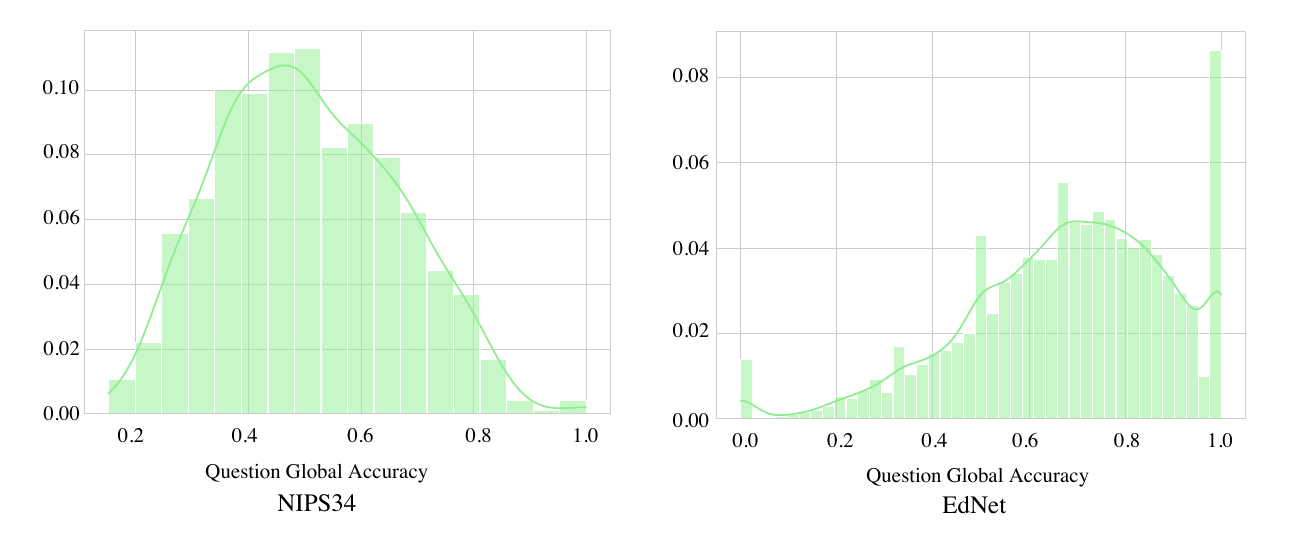}
    \caption{The distribution of question global accuracy in NIPS34 and EdNet Datasets.}
    \label{fig:question_correct_distribution}
\end{figure}
To mitigate the risk of less interactive questions getting overfitted, we leverage the power of the contrastive learning technique to enhance the representation of the less interactive question, thereby improving the accuracy of its corresponding question knowledge acquisition state.
Contrastive learning focuses on bringing positive pair samples closer together while pushing negative pair samples further apart. 
To achieve this, we define two major procedures: (1) construction of positive and negative pairs, and (2) contrastive loss.

\subsubsection{Construction of Positive and Negative Pairs}
Different from the approach taken by \citet{cl4kt}, where positive and negative pairs are obtained through coarse automatic data augmentation, we have adopted a more fine-grained approach to design positive and negative pairs for less interactive questions.
Specifically, we designate the less interactive questions as the anchor, denoted as $q^{\text{anchor}}_{i}$ ($q^{\text{anchor}}_{i} \in \mathcal{Q}_{less} = \{q^{\text{anchor}}_{0}, ...q^{\text{anchor}}_{i}$\}).
To identify its positive question, we first find the positive question candidate $p^{*\text{anchor}}_{j}$ ($p^{*\text{anchor}}_{j} \in \mathcal{P}^{*}_{high} = \{p^{*\text{anchor}}_{0}, ...p^{*\text{anchor}}_{j}\}$), which is the relative highly interactive question\footnote{We use the interaction frequency threshold $\epsilon$ to distinguish less interactive and relative highly interactive questions.} and has the same knowledge concepts as $q^{\text{anchor}}_{i}$.
Subsequently, we filter out the question from the pool of positive question candidates $\mathcal{P}^{*}_{high}$ that shares a similar difficulty level to $q^{\text{anchor}}_{i}$.
The filtered-out question $p^{\text{anchor}}_{k}$ ($p^{\text{anchor}}_{k} \in \mathcal{P}_{high} = \{p^{\text{anchor}}_{0}, ...p^{\text{anchor}}_{k}\}$) is the final positive question that encompasses the fine-grained semantic alignment information.
To facilitate the judgment of whether questions have a similar difficulty level, we begin by calculating the question's global accuracy. 
This is achieved by dividing the number of correct responses to a question by its total interactions across all students. 
Subsequently, we categorize questions into different classes based on the distribution of their global accuracy. 
Questions within the same difficulty level class are considered to have a similar difficulty level.
The categorization process satisfies two criteria: (1) each difficulty level class should contain a number of questions to provide a certain quantity of positive samples, and (2) the range of global question accuracy within each difficulty level class should not exhibit a wide span. 
Taking NIPS34 and EdNet datasets as examples (as shown in Figure \ref{fig:question_correct_distribution}), we categorize the questions in NIPS34 dataset into six difficulty level classes: Level A ([0,0.3]), Level B ((0.3, 0.4]), Level C ((0.4, 0.5]), Level D ((0.5, 0.6]), Level E ((0.6, 0.7]) and Level F ((0.7, 1]).

For selecting negative questions, the most straightforward approach is to choose the questions with different concepts compared to the anchor question $q^{\text{anchor}}_{i}$. 
However, the negative questions obtained through this method can be considered as easy negative samples, which do not provide sufficient negative information for the contrastive learning process.
To address this limitation, we adopt a strategy to obtain negative questions at a harder level. 
Specifically, we utilize the positive question candidates $\mathcal{P^{*}}_{high}$, which are obtained in the first step of selecting positive questions. 
From these positive question candidates, we select the question that belongs to different difficulty level classes than $q^{\text{anchor}}_{i}$ as negative question $n^{\text{anchor}}_{l}$ ($n^{\text{anchor}}_{l} \in \mathcal{N}_{high} = \{n^{\text{anchor}}_{0}, ...n^{\text{anchor}}_{l}\}$). 
To mitigate the errors introduced by the procedure of categorization of question difficulty level, we ensure that the selected difficulty level is at least two classes apart from the difficulty level of $q^{\text{anchor}}_{i}$. 
For example, in the NIPS34 dataset, the negative difficulty level classes for Level A are Levels C, D, E, and F.
We argue that the positive and negative pairs selected by our proposed method can bring more fine-grained information during the contrastive learning process.

\begin{figure}[!tbph]
    \centering
    \includegraphics[width=0.7\columnwidth]{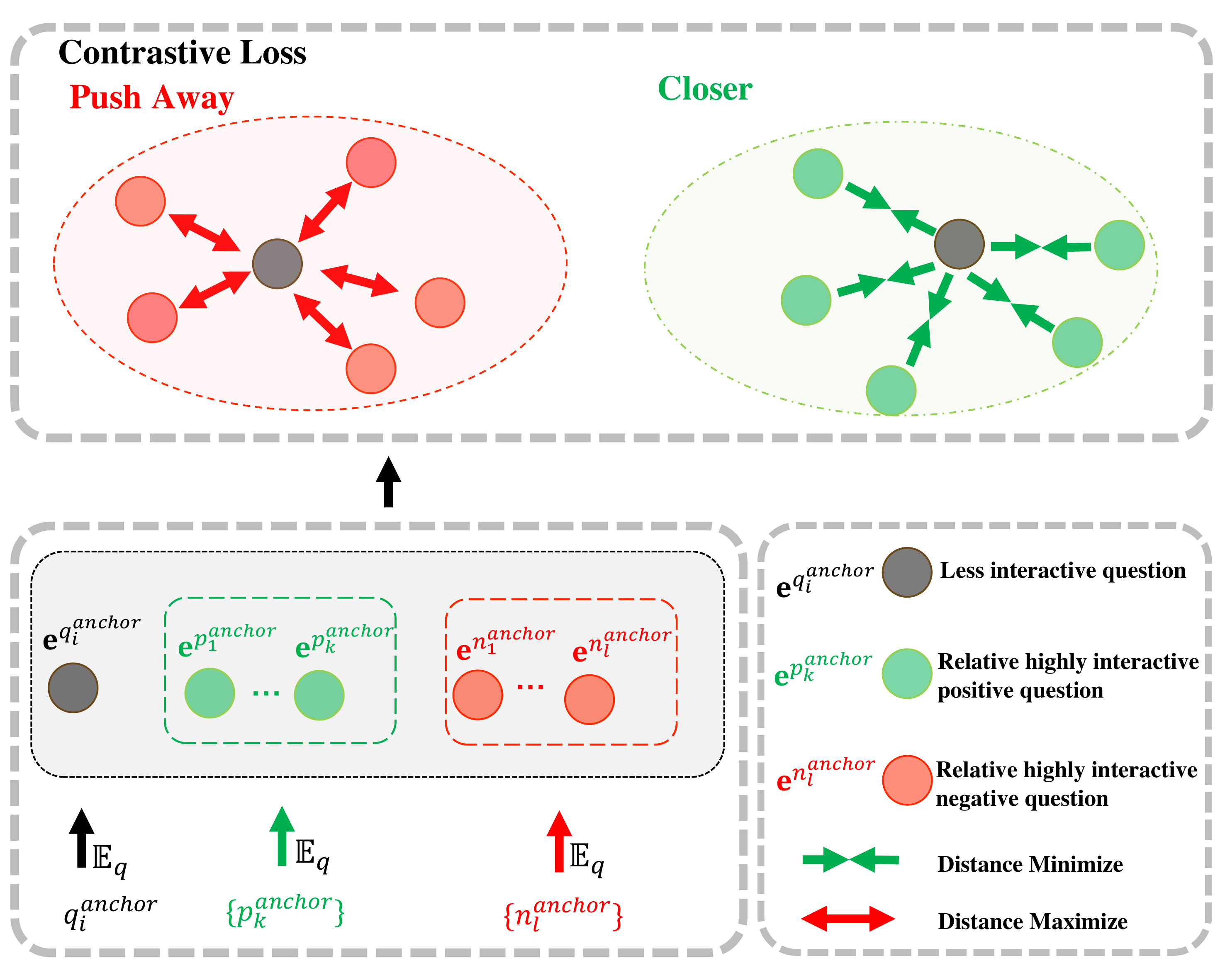}
    \caption{A graphical illustration of the fine-grained question-centric contrastive learning framework.}
    \label{fig:contrastive}
\end{figure}
\subsubsection{Contrastive Loss}
\label{sec:contrastive_loss}
To better adapt the contrastive loss to different datasets, we carefully design the contrastive loss to enhance the representations of less interactive questions $\mathcal{Q}_{less}$.
Specifically, we first extract the embedding of $q^{\text{anchor}}_{i}$, $p^{\text{anchor}}_{k}$, and $n^{\text{anchor}}_{l}$ from the question embedding matrix $\mathbf{Q}$ and its copy version $\mathbf{Q}_c$\footnote{We borrow the idea in \cite{momentum} to enhance the stability of the training process by duplicating another question embedding matrix at each epoch to encode positive and negative samples.} according to their question index:
\begin{equation}
\mathbf{e}^{q^{\text{anchor}}_{i}}, \mathbf{e}^{p^{\text{anchor}}_{k}}, \mathbf{e}^{n^{\text{anchor}}_{l}} = \mathbf{Q}(q^{\text{anchor}}_{i}), \mathbf{Q}_{c}(p^{\text{anchor}}_{k}, n^{\text{anchor}}_{l})
\end{equation}

\noindent where $\mathbf{e}^{q^{\text{anchor}}_{i}}, \mathbf{e}^{p^{\text{anchor}}_{k}}$, and $\mathbf{e}^{n^{\text{anchor}}_{l}}$ represent the embedding of $q^{\text{anchor}}_{i}, p^{\text{anchor}}_{k}$, and $n^{\text{anchor}}_{l}$.

To pull positive questions close to $q^{\text{anchor}}_{i}$ and push negative questions away from $q^{\text{anchor}}_{i}$ in a controlled way, we design a distance-aware contrastive loss. 
Mathematically, the contrastive loss is formulated as:

\begin{equation}
\begin{split}
\label{eq:contrastive_loss}
\mathcal{L}_{\mbox{{\tiny CL}}} &= \frac{1}{I}\sum_{i=1}^{I}(\mathcal{L}^{i}_{p} + \mathcal{L}^{i}_{n} ) / \tau \\
\mathcal{L}^{i}_{p} &= \frac{1}{K}\sum^{K}_{k=1}{\text{{ReLU}}(d^{p}_{k} - \sigma_p)} \\
\mathcal{L}^{i}_{n} &= \frac{1}{L}\sum^{L}_{l=1}{\text{{ReLU}}(d^{n}_{l} - \sigma_n)} \\
d^{p}_{k} &= \|\mathbf{e}^{q^{\text{anchor}}_{i}}-\mathbf{e}^{p^{\text{anchor}}_{k}}\|  \\
d^{n}_{l} &= \|\mathbf{e}^{q^{\text{anchor}}_{i}}-\mathbf{e}^{n^{\text{anchor}}_{l}}\| \\
\end{split}
\end{equation}

\noindent where $d^{p}_{k}$ is the distance of positive pair ($q^{\text{anchor}}_{i}$, $p^{\text{anchor}}_{k}$), $d^{n}_{l}$ is the distance of negative pair ($q^{\text{anchor}}_{i}$, $n^{\text{anchor}}_{l}$). $\sigma_p$ and $\sigma_n$ are the positive and negative distance margins, which can control the extent of representation learning. 
$I$, $K$, and $L$ represent the number of less interactive questions, its positive questions and its negative questions respectively.
$\tau$ is the temperature to scale the contrastive loss. 
The overview of our proposed fine-grained question-centric contrastive learning is illustrated in Figure~\ref{fig:contrastive}.

\subsection{Interpretable Prediction Layer}
\label{sec:prediction_layer}
Explaining the decision-making process of DLKT models is generally challenging. Therefore, we introduce an IRT-based prediction layer that leverages the estimated acquisition states obtained from Sections~\ref{sec:MQKA} and~\ref{sec:MCKA} to derive the final prediction result. This is in contrast to directly generating predictions based on the hidden states of neural networks.
Specifically, we first extract the question knowledge acquisition score $\alpha_{t+1}$ and the concept knowledge acquisition score $\bar{\beta}_{t+1}$ from the question knowledge acquisition state $\mathbf{\gamma}^q_{t}$ and concept knowledge acquisition state respectively $\mathbf{\gamma}^c_{t}$:

\begin{equation}
    \begin{split}
        \alpha_{t+1} &= \mathbf{\gamma}^q_{t} \cdot \delta(q_{t+1}) \\
        \bar{\beta}_{t+1} &= \frac{1}{m} \sum_{m = 1}^m \mathbf{\gamma}^c_{t} \cdot \delta(k^{m}_{t+1}) * \mathbb{I}(k^m_{t+1} \in \{k\}_{t+1})\\
    \end{split}
\end{equation}
\noindent where $m$ is the number of KCs in question $q_{t+1}$, $\delta(q_{t+1})$ and $\delta(k^{m}_{t+1})$ is the one-hot encoded vector of next question and its corresponding m-th KC.

Then, similar to previous work by \citet{yeung2019deep}, we use the IRT function to calculate the probability of a correct answer.
We let IRT function directly take the linear combined scores of question knowledge acquisition score\footnote{The question knowledge acquisition score can serve as an indicator of both the student's acquisition of the question and the difficulty level of the question itself, as when a student obtains a high acquisition score for a specific question, it signifies that the difficulty level of that question has reduced for the student.} and the concept knowledge acquisition score to enhance the interpretability of prediction process i.e.:

\begin{equation} 
\hat{r}_{t+1} = \sigma(\alpha_{t+1} + \bar{\beta}_{t+1})
\end{equation}
We explicitly choose not to include any learnable parameters inside the IRT-based prediction function for better interpretability.
This approach enables educators to comprehend how predictions are obtained from our Q-MCKT, facilitating them to accept the rationale behind the results and apply them in designing teaching activities and personalized learning strategies for students.
For instance, leveraging the concept knowledge acquisition state of students, educators can identify the specific areas where students may face challenges.
This allows for a focused effort on reinforcing these concepts through a set of relevant practice from the question bank.
To better conduct personalized teaching, educators can further tailor the relevant practices based on student's estimated question acquisition scores, i.e., question knowledge acquisition state.
This involves suggesting practice questions in a graduated difficulty sequence, starting from easier questions and progressing towards more challenging ones.
Note that the Q-MCKT framework is designed to enhance the interpretability of prediction results, differing from the interpretability of model parameters.
The overview of our prediction process is shown in Figure \ref{fig:overview}.

\subsection{Optimization of Q-MCKT}
\label{sec:train}
Our Q-MCKT model is optimized by minimizing the binary cross entropy loss $\mathcal{L}_{\mbox{\tiny KT}}$ between the ground-truth responses $r_i$s and the estimated probabilities $\hat{r}_i$s from the IRT layer and the contrastive loss $\mathcal{L}_{\mbox{\tiny CL}}$ in Section~\ref{sec:contrastive_loss}.
Furthermore, to directly improve the modeling ability of MQKA and MCKA modules, we explicitly cast question knowledge acquisition scores $\alpha_{t}s$ and concept knowledge acquisition scores $\bar{\beta}_{t}s$ via sigmoid function and add as the optimization terms into the overall model training process. 
Therefore, the final optimization function is:
\begin{equation}
\mathcal{L} = \mathcal{L}_{\mbox{\tiny KT}} + \lambda_{1}\bigl(\mathcal{L}_{*}(\boldsymbol{\alpha}) + \mathcal{L}_{*}(\boldsymbol{\bar{\beta}}) \bigl) + \lambda_{2}\mathcal{L}_{\mbox{\tiny CL}}
\end{equation}

\noindent where $\lambda_1$ and $\lambda_2$ are the hyper-parameters. 
$\boldsymbol{\alpha}$ denotes the collection of question knowledge acquisition scores ($\boldsymbol{\alpha} = \{\alpha_i\}$) and $\boldsymbol{\bar{\beta}}$ denotes the collection of concept knowledge acquisition scores ($\boldsymbol{\bar{\beta}} = \{\bar{\beta_i}\}$).

$\mathcal{L}_{\mbox{\tiny KT}}$ and $\mathcal{L}_{*}(\mathbf{z})$ are defined as follows: 

\begin{equation*}
\mathcal{L}_{\mbox{\tiny KT}} = - \sum_{i} \bigl( r_i \log \hat{r}_i + (1-r_i) \log (1-\hat{r}_i) \bigl) 
\end{equation*}

\begin{equation*}
\mathcal{L}_{*}(\mathbf{z})  = - \sum_i \bigl( r_i \log \sigma(z_i) + (1-r_i) \log (1-\sigma(z_i)) \bigl)  
\end{equation*}

\noindent where $\mathbf{z}$ can be either $\boldsymbol{\alpha}$ or $\boldsymbol{\bar{\beta}}$, $\mathcal{L}_{*}$ is a binary cross entropy loss function.

\section{Experiments}
\label{sec:exp}
In this section, we present details of our experiment settings and the corresponding results. 
We conduct comprehensive analysis and investigations to illustrate the effectiveness of our Q-MCKT model.
Specifically, we aim to answer the following research questions: 
(\textbf{RQ1}) How does the proposed Q-MCKT framework perform compared to the state-of-the-art KT methods? 
(\textbf{RQ2}) What is the influence of components in Q-MCKT, such as the interpretable prediction layer and concept knowledge acquisition module? 
(\textbf{RQ3}) How does the CL affect the performance of Q-MCKT?
(\textbf{RQ4}) How does the number of experts influence the performance of Q-MCKT?
(\textbf{RQ5}) Can Q-MCKT model accurately estimate the knowledge acquisition state?
(\textbf{RQ6}) How does Q-MCKT produce interpretable prediction results?

\subsection{Datasets}
\label{sec:dataset}
\begin{table*}[!hptb]
    \small
    \centering
    \caption{Data statistics of four datasets. \# of Ss/Is/Qs denote the number of students, interactions and questions. \emph{Avg. per Q} and \emph{Avg. per KC} denotes the number of KCs per question and the number of questions per KC. \emph{Avg. Q} denotes the average interactions per question.} 
    \begin{tabular}{lccccccc}
    \toprule
    Dataset & \# of Ss & \# of Is & \# of Qs & \# of KCs & Avg. per Q  & Avg. per KC  & Avg. Q \\ \hline
    \textbf{ASSIST2009}  & 4,217    & 346,860      & 26,688    & 123 & 1.197   & 216.9  & 12.9      \\
    \textbf{Algebra2005} & 574     & 809,694      & 210,710   & 112 & 1.364   & 1881.3   & 3.8    \\
    \textbf{NIPS34}      & 4,918    & 1,382,727     & 948      & 57  & 1.015   & 16.6    & 1458.6     \\
    \textbf{EdNet}       & 4,999    & 597,042     & 11,901     & 188 & 2.245    & 63.3   & 50.2 \\
    \bottomrule
    \end{tabular}
    \label{tab:sta}
\end{table*}

We use four widely used and publicly available datasets to evaluate the performance of Q-MCKT model:

\begin{itemize}[leftmargin=*]
    \item \textbf{ASSISTments2009\footnote{https://sites.google.com/site/assistmentsdata/home/2009-2010-assistment-data/skill-builder-data-2009-2010} (ASSIST2009)}: is collected from ASSISTment online tutoring platform in the school year 2012-2013 that students are assigned to answer similar exercises from the skill builder problem sets. 
    \item \textbf{Algebra 2005-2006\footnote{https://pslcdatashop.web.cmu.edu/KDDCup/} (Algebra2005)}: stems from KDD Cup 2010 EDM Challenge which includes 13-14 year-old students’ interactions with Algebra questions. It has detailed step-level student responses to the mathematical problems.
    \item \textbf{NeurIPS2020 Education Challenge\footnote{https://eedi.com/projects/neurips-education-challenge} (NIPS34)}: is released in Task 3 and Task 4 of NeurIPS2020 Education Challenge, it contains students’ answers to mathematics questions from Eedi which millions of students interact with daily around the globe. \cite{wang2020instructions}.
    \item \textbf{EdNet Dataset\footnote{https://company.riiid.co/en/product} (EdNet)}: is collected over two years from the intelligent online tutoring platform named \textit{Riid TUTOR} dedicated to practicing English for international communication (TOEIC) assessment in South Korea \cite{choi2020ednet}.
\end{itemize}
To conduct reproducible experiments, we rigorously follow the data pre-processing steps suggested in \cite{liu2022pykt}. 
We remove student sequences shorter than 3 attempts. 
Data statistics are summarized in Table \ref{tab:sta}.

\subsection{Baselines}
\label{sec:baselines}
We compare our Q-MCKT with the following state-of-the-art DLKT models to evaluate the effectiveness of our approach:
\begin{itemize}[leftmargin=*]
    \item \textbf{DKT}: leverages an LSTM layer to encode the student knowledge state to predict the students' performances \cite{dkt}.
    \item \textbf{DKT+}: an improved version of DKT to solve the reconstruction and non-consistent prediction problems \cite{dkt+}.
    \item \textbf{DKT-F}: an extension of DKT that model the students' forgetting behavior on the DKT model \cite{nagatani2019augmenting}.
    \item \textbf{DKVMN}: designs a static key matrix to store the relations between the different KCs and a dynamic value matrix to update the students' knowledge state \cite{zhang2017dynamic}.
    \item \textbf{ATKT}: performs adversarial perturbations into student interaction sequence to improve model's generalization ability \cite{Adversarial-KT}.
    \item \textbf{GKT}: utilizes the graph structure to predict the students' performance \cite{nakagawa2019graph}.
    \item \textbf{SAKT}: uses self-attention to capture relations between exercises and student responses \cite{sakt}.
    \item \textbf{SAINT}: uses the Transformer-based layer to capture students' exercise and response sequences \cite{choi2020towards}.
    \item \textbf{AKT}: employ the Rasch model to get a series of Rasch model-based embeddings, which can capture individual differences among responses \cite{AKT}.
    \item \textbf{SKVMN}: designs a Sequential Key-Value Memory Networks to capture long-term dependencies in an exercise sequence \cite{abdelrahman2019knowledge}.
    \item \textbf{DeepIRT}: a combination of the IRT and DKVMN to enhance the interpretability of memory augmented models \cite{yeung2019deep}.
    \item \textbf{qDKT}: predicts the future performance of student knowledge state at the question level \cite{qdkt}.
    \item \textbf{AT-DKT}: improves the prediction performance of the original deep knowledge tracing model with two auxiliary learning tasks \cite{AT-DKT}.
    \item \textbf{SimpleKT}: leverages Rasch model in psychometrics to provide a strong but simple baseline method to deal with the KT task \cite{liu2023simplekt}.
\end{itemize}

\subsection{Experimental Setup}
\label{sec:exp_setting}
To evaluate the model's performance, we perform 5-fold cross-validation for every combination of models and datasets.
We set the maximum length of model input sequence to 200. 
We use 80\% of student sequences for training and validation, and use the rest 20\% of student sequences for model evaluation. 
We adopt Adam optimizer \cite{kingma2015adam} to train all the models. 
The number of training epochs is set to 200. 
We choose to use an early stopping strategy that stops optimization when the AUC score is failed to get the improvement on the validation set in the latest 10 epochs. 
We adopt the Bayesian search method to find the best hyper-parameters for each fold by using wandb tool\footnote{https://wandb.ai/}.
Specifically, the hyper-parameter $\lambda_1$ and $\lambda_2$, the learning rate and the embedding size $d$ are searching from \{0, 0.5, 1, 1.5, 2\}, \{1e-3, 1e-4, 1e-5\}, \{64, 256\} respectively. 
Following all existing DLKT research \cite{AT-DKT,Adversarial-KT,AKT,choi2020towards}, we use the Area Under the Curve (AUC) as the main evaluation metric.
We also choose to use Accuracy as the secondary evaluation metric.

\subsection{Overall Performance (RQ1)}
\label{sec:overall_performance}
\vspace{0.5cm}
\begin{table}[!bpht]
\small
\caption{The AUC and Accuracy performance of all the baseline models and our proposed Q-MCKT model.
We highlight the highest results with bold.
Marker $*$, $\circ$, and $\bullet$ indicates whether our proposed model is statistically superior/equal/inferior to the compared baseline models (using paired t-test at 0.01 significance level). 
}
\label{tab:overall}
\begin{center}
\setlength\tabcolsep{0.5pt}
\begin{tabular}{lccccccccc}
\hline
                                                   & \multicolumn{4}{c}{AUC}                        & \multicolumn{1}{c}{} & \multicolumn{4}{c}{Accuracy}                \\ \cline{2-5} \cline{7-10}
\multirow{-2}{*}{Model}                            & ASSIST2009 & Algebra2005 & NIPS34   & EdNet  &                   & ASSIST2009   & Algebra2005         & NIPS34   & EdNet      \\ \hline
\textbf{DKT}                                       & 0.7541±0.0011*  & 0.8149±0.0011*        & 0.7689±0.0002*  & 0.6133±0.0006* &                      & 0.7244±0.0014*       & 0.8097±0.0005* & 0.7032±0.0004* & 0.6462±0.0028*     \\
\textbf{DKT+}                                      & 0.7547±0.0017*  & 0.8156±0.0011*        & 0.7696±0.0002*  & 0.6189±0.0012* &                      & 0.7248±0.0009*       & 0.8097±0.0007* & 0.7076±0.0002* & 0.6402±0.0021*     \\
\textbf{DKT-F}                                     & -        & 0.8147±0.0013*        & 0.7733±0.0003*  & 0.6168±0.0019* &                      & -             & 0.8090±0.0005* & 0.7076±0.0002* & 0.6402±0.0021*     \\
\textbf{ATKT}                                   & 0.7470±0.0008*  &0.7995±0.0023*        & 0.7665±0.0001*  & 0.6065±0.0003* &                      & 0.7208±0.0009*       & 0.7998±0.0019*         & 0.7013±0.0002* & 0.6369±0.0009*    \\
\textbf{GKT}                                       & 0.7424±0.0021*  & 0.8110±0.0009*        & 0.7689±0.0024*  & 0.6223±0.0017* &                      & 0.7153±0.0032*       & 0.8088±0.0008*         & 0.7014±0.0028* & 0.6625±0.0064*     \\
\textbf{SAKT}                                      & 0.7246±0.0017*  & 0.7880±0.0063*        & 0.7517±0.0005*  & 0.6072±0.0018* &                      & 0.7063±0.0018* & 0.7954±0.0020*         & 0.6879±0.0004* & 0.6391±0.0041*     \\
\textbf{SAINT}                                     & 0.6958±0.0023*  & 0.7775±0.0017*        & 0.7873±0.0007*  & 0.6614±0.0019* &                      & 0.6936±0.0034*       & 0.7791±0.0016*         & 0.7180±0.0006* & 0.6522±0.0024*     \\
\textbf{AKT}                                       & 0.7853±0.0017*   & 0.8306±0.0019*        & \pmb{0.8033±0.0003}$\circ$   & 0.6721±0.0022*  &                      & \pmb{0.7392±0.0021}$\bullet$        & 0.8124±0.0011*         & \pmb{0.7323±0.0005}$\circ$  & 0.6655±0.0042*   \\
\textbf{SKVMN}                                     & 0.7332±0.0009*   & 0.7463±0.0022*        & 0.7513±0.0005*   & 0.6182±0.0114*  &                      & 0.7156±0.0012*        & 0.7837±0.0023*         & 0.6885±0.0005*  & 0.6555±0.0152*   \\
\textbf{DKVMN}                                     & 0.7473±0.0006*  & 0.8054±0.0011*  & 0.7673±0.0004*  & 0.6158±0.0022* &                      & 0.7199±0.0010*       & 0.8027±0.0007*         & 0.7016±0.0005* & 0.6444±0.0030*     \\
\textbf{DeepIRT}                                   & 0.7465±0.0006*   & 0.8040±0.0013*        & 0.7672±0.0006*   & 0.6173±0.0008*  &                      & 0.7195±0.0004*        & 0.8037±0.0009*         & 0.7014±0.0008*  & 0.6457±0.0033*   \\
\textbf{qDKT}                                      & 0.7332±0.0009*  & 0.7485±0.0017*        & 0.7995±0.0008*   & 0.6987±0.0010*  &                       & 0.6787±0.0039*       & 0.7262±0.0012*         & 0.7299±0.0007*  & 0.6922±0.0004*   \\
\textbf{AT-DKT}                                    & 0.7555±0.0005*   & 0.8246±0.0019*        & 0.7816±0.0002*   & 0.6249±0.0020*  &                      & 0.7250±0.0007*        & 0.8144±0.0008*         & 0.7146±0.0002*  & 0.6512±0.0039*   \\
\textbf{SimpleKT}                                  & 0.7744±0.0018*   & 0.8254±0.0003*        & \pmb{0.8035±0.0000}$\circ$   & 0.6599±0.0027*  &                      & 0.7320±0.0012*        & 0.8083±0.0005*         & \pmb{0.7328±0.0001}$\circ$  & 0.6557±0.0029*   \\
\textbf{Q-MCKT}                             & \pmb{0.7876±0.0016}   & \pmb{0.8429±0.0011}       & \pmb{0.8035±0.0004} & \pmb{0.7228±0.0023}  &                      & 0.7376±0.0011        & \pmb{0.8239±0.0006}         & \pmb{0.7329±0.0005}  & \pmb{0.7067±0.0008} \\ \hline

\end{tabular}
\end{center}
\end{table}

The overall model performance is reported in Table \ref{tab:overall}. 
From Table \ref{tab:overall}, we make the following observations: (1) Our proposed model significantly outperforms 14 baselines on all four datasets in terms of AUC metric, and only has one loss with AKT in ASSIST2009 dataset in terms of Accuracy metric.
More importantly, as a representative of the deep sequential KT models, compared with DKT, our proposed model improves the AUC by 3.30\%, 2.80\%, 3.46\%, and 10.95\% on four datasets. 
That shows our proposed method can significantly improve the performance. 
(2) When comparing performance on ASSIST2009, Algebra2005 to NIPS34 and EdNet, DLKT models behave quite differently. 
For example, DKT significantly outperforms qDKT on the ASSIST2009 and Algebra2005 datasets in AUC by 5.30\% and 6.60\% but is beaten by qDKT by 3.10\% and 8.54\% on the NIPS34 and EdNet datasets. 
Meanwhile, SAINT performs terrible in ASSIST2009 and Algebra2005 datasets, but is good on NIPS34 and EdNet datasets. 
We argue that it is because the average number of interactions per question (Avg .Q) for the ASSIST2009 and Algebra2005 are smaller than that for the NIPS34 and EdNet datasets (As we can see from Table \ref{tab:sta}, the values of Avg. Q are 12.9 and 3.8 in ASSIST2009 and Algebra2005 datasets). 
This results in the question-centric modeling overfitting to the data, which adversely affects predictive accuracy. 
(3) Results between DeepIRT and DKVMN are very close on three datasets, which empirically shows that the IRT function won't sacrifice the model prediction ability too much. 
(4) AKT is a very strong baseline, it outperforms almost all other baseline methods on all datasets. 
Since it learns the Rasch model-based embeddings that implicitly model question difficulties, it further empirically verifies the importance of considering question representations when building the DLKT models.
(5) Our proposed Q-MCKT exhibits a more substantial improvement over the baselines on the EdNet dataset compared to the other three datasets. 
We attribute this to the fact that the EdNet dataset has a larger number of average KCs per question (Avg. per Q = 2.245), indicating that the majority of questions in EdNet encompass more than two KCs.
However, most baseline methods encode the student's learning sequence at the concept level, utilizing concepts to index questions. 
In this process, each question will be transformed into several KC interactions, and questions covering the same KC are treated as identical questions. 
This approach disregards much valuable question-centric information and introduces noise in the encoding process.
This is particularly pronounced in datasets with larger Avg. per Q, such as EdNet, where questions containing more KCs result in a transformed process that includes additional noise.
In contrast, as illustrated in Section~\ref{sec:interaction_projection}, our Q-MCKT adopts a different approach by encoding the student's learning sequence at the question level. 
This allows for the capture of both question- and concept-centric information in a more granular manner. 
Consequently, this approach effectively mitigates the aforementioned problems associated with datasets featuring larger Avg. per Q, as exemplified by EdNet.

\subsection{Ablation Study (RQ2)}
\label{sec:abl_study}

\begin{table}[!hptb]
    \small
    \caption{The AUC and Accuracy performance of different variants in our proposed Q-MCKT. We highlight the highest results with bold. Marker $*$, $\circ$ and $\bullet$ indicates whether our Q-MCKT model is statistically superior/equal/inferior to its different variants (using paired t-test at 0.01 significance level).}
    \setlength\tabcolsep{0.5pt}
    \centering
    \begin{tabular}{lccccccccc}
    \toprule
            & \multicolumn{4}{c}{AUC}                        & \multicolumn{1}{c}{} & \multicolumn{4}{c}{Accuracy}                \\ \cline{2-5} \cline{7-10}
            \multirow{-2}{*}{Method} & ASSIST2009  & Algebra2005 & NIPS34 & EdNet & & ASSIST2009 & Algebra2005  & NIPS34 & EdNet          \\ 
            \hline
            Q-MCKT   & \pmb{0.7876±0.0016}   & \pmb{0.8429±0.0011}    & \pmb{0.8035±0.0004} & \pmb{0.7228±0.0023} &    &\pmb{0.7376±0.0011}      & \pmb{0.8239±0.0006}         & \pmb{0.7329±0.0005}  & \pmb{0.7067±0.0008}            \\ \hline
             w/o MCKA           & 0.7328±0.0031*        & 0.7573±0.0007*              &0.8023±0.0002* & 0.7112±0.0019*  &   & 0.7010±0.0020*        & 0.7310±0.0007*              &0.7316±0.0002* & 0.7006±0.0005*                \\
             w/o MQKA           & 0.7599±0.0008*        & 0.8277±0.0006*              &0.7686±0.0007* & 0.6539±0.0015*  &   & 0.7287±0.0006*        & 0.8171±0.0004*              &0.7030±0.0006* & 0.6789±0.0005*                 \\
             w/o MoE            & 0.7822±0.0017*        & 0.8361±0.0008*       & 0.8014±0.0002*           & 0.7211±0.0020*  & & 0.7332±0.0014*        & 0.8204±0.0006*       & 0.7314±0.0005*           & 0.7049±0.0014*       \\
             w/o CL           & 0.7865±0.0014*        & 0.8421±0.0007*              &0.8004±0.0002* & 0.7207±0.0022*      &   & 0.7354±0.0012*        & 0.8232±0.0007*              &0.7302±0.0004* & 0.7045±0.0011*                 \\
             w/o IRT            & 0.7847±0.0019*        & 0.8313±0.0006*         & 0.7976±0.0003*                  & 0.7214±0.0031*    &      & 0.7351±0.0008*        & 0.8227±0.0020*         & 0.7286±0.0003*                  & 0.7038±0.0027*    \\
            \bottomrule
    \end{tabular}
    \label{tab:ab_study}
\end{table}

We examine the effect of key components by constructing five model variants in Table \ref{tab:ab_study}. 
``w/o'' means excludes such module from Q-MCKT.
From Table \ref{tab:ab_study}, we can observe that (1) comparing Q-MCKT and Q-MCKT w/o MCKA, the performance drops a lot in ASSIST2009 (5.48\% in AUC) and Algebra2005 (8.56\% in AUC) datasets.
We believe this is due to the low values of Avg. Q in ASSIST2009 and Algebra2005 datasets (shown in Table~\ref{tab:sta}).
Therefore, the acquisition of general concept knowledge provided by the MCKA module becomes crucial in such circumstances.
(2) Comparing Q-MCKT and Q-MCKT w/o MQKA, the overall performance is degraded.
This indicates that it is important to incorporate question information in DLKT models.
(3) Having multiple experts leads to better performance compared to having only one expert across all datasets.
This finding demonstrates the effectiveness of our proposed multi-experts knowledge acquisition modules in providing a more robust and accurate knowledge acquisition state for the interpretable prediction layer.
Another interesting finding is that the magnitude of improvement differs across datasets. 
The Algebra2005 and ASSIST2009 datasets show an improvement of 0.54\% and 0.68\% in AUC, while the NIPS34 and EdNet datasets exhibit smaller improvements of 0.21\% and 0.17\% in AUC respectively. 
This indicates that the mixture of experts technique is more effective for datasets with a low value of Avg. Q, as MQKA can provide a more robust question knowledge acquisition state, which mitigates the negative impact caused by the potential overfitted question embedding.
(4) Incorporating contrastive learning as an auxiliary task can be beneficial for datasets with larger average interactions per question (Avg. Q) values, because the relative highly interactive questions are well-represented in this context, providing sufficient supplemental information for the corresponding less interactive questions. In contrast, datasets with relatively low Avg. Q value show limited improvement (e.g. an improvement in AUC of 0.31\% in NIPS34, while a smaller improvement in AUC of 0.05\% in Algebra2005).
(5) Comparing Q-MCKT and Q-MCKT w/o IRT, the presence of an IRT-based interpretable prediction layer not only enhances interpretability but also improves prediction performance.
This finding highlights the dual benefits of incorporating IRT into the DLKT models, paving the way for more interpretable and accurate educational assessments.

\subsection{Analysis of Contrastive Learning (RQ3)}
\label{sec:impact_cl}
\begin{table}[!hptb]
    \centering
    \caption{The AUC and Accuracy performance of Q-MCKT and its w/o CL variant on less interactive questions. We highlight the highest results with bold. Marker $*$, $\circ$, and $\bullet$ indicate whether our proposed model is statistically superior/equal/inferior to the variant (using paired t-test at 0.01 significance level).}
    \setlength\tabcolsep{4pt}
    \begin{tabular}{lccccccc}
    \toprule
            & \multicolumn{3}{c}{AUC}       & \multicolumn{1}{c}{}  & \multicolumn{3}{c}{Accuracy}                \\ \cline{2-4} \cline{6-8}
    \multirow{-2}{*}{Method}    & ASSIST2009               & NIPS34             & EdNet &    & ASSIST2009               & NIPS34             & EdNet    \\
    \hline
    Q-MCKT    & \pmb{0.7789±0.0019}          & \pmb{0.7805±0.0007}           & \pmb{0.7098±0.0024}  & & \pmb{0.7354±0.0021}          & \pmb{0.7213±0.0004}           & \pmb{0.7032±0.0012}      \\ 
    w/o CL    & 0.7765±0.0016*          & 0.7716±0.0005*             & 0.7027±0.0015*    & & 0.7339±0.0017*          & 0.7158±0.0005*             & 0.7001±0.0014*            \\
    \bottomrule
    \end{tabular}
    \label{tab:cl_impact}
\end{table}

\begin{figure}[!h]
    \centering
    \includegraphics[width=0.55\columnwidth]{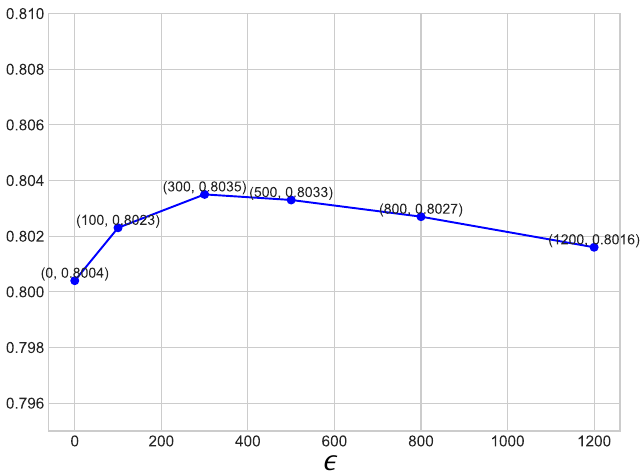}
    \caption{The impact of $\epsilon$ on overall performance in NIPS34 dataset.}
    \label{fig:episilon}
\end{figure}

To further investigate the effectiveness of our proposed contrastive learning procedure, we conduct additional experiments to analyze the impact of contrastive learning specifically on less interactive questions\footnote{To save computation, we only experiment on the three datasets, as the performance with and without CL in Algebra2005 dataset is comparatively less significant when compared to the other three datasets.}. 
As presented in Table \ref{tab:cl_impact}, the results demonstrate that our proposed contrastive learning procedure helps the less interactive question learn a more general representation under relative rare updated times.
This more general representation leads to more robust and accurate question knowledge acquisition state for the prediction layer to get a better performance (e.g., an improvement in AUC of 0.89\% in NIPS34).

We also conduct experiments to examine the influence in the construction of positive and negative questions.
As illustrated in Figure~\ref{fig:episilon}, when the value of $\epsilon$ increases, the performance initially improves but then deteriorates. 
This can be attributed to the fact that when $\epsilon$ exceeds a certain threshold, there are insufficient relative highly interactive questions available to serve as positive and negative examples for less interactive questions. 
As a result, the representation learning of the less interactive question is hindered due to a lack of informative supplemental information.

\subsection{Impact of Expert Number (RQ4)}
\label{sec:impact_moe}

\begin{table}[!hptb]
    \small
    \caption{The AUC and Accuracy performance of different numbers of experts in Q-MCKT. We highlight the highest results with bold.}
    \centering
    \setlength\tabcolsep{2.1pt}
    \begin{tabular}{l|ccccccccc}
    \toprule
                  & \multicolumn{4}{c}{AUC}                        & \multicolumn{1}{c}{} & \multicolumn{4}{c}{Accuracy}                \\ \cline{2-5} \cline{7-10}
            \multirow{-2}{*}{e}    & ASSIST2009            & Algebra2005          & NIPS34             & EdNet  & & ASSIST2009            & Algebra2005          & NIPS34             & EdNet          \\
             \hline
             1    & 0.7822±0.0017        & 0.8361±0.0008       & 0.8014±0.0002           & 0.7211±0.0020     &   & 0.7332±0.0014        & 0.8204±0.0006       & 0.7314±0.0005           & 0.7049±0.0014  \\ 
             2    & 0.7864±0.0014        & \pmb{0.8429±0.0011}  & \pmb{0.8035±0.0004}      & 0.7221±0.0022     &   & 0.7365±0.0009        & \pmb{0.8239±0.0006}  & \pmb{0.7329±0.0005}      & 0.7058±0.0015             \\
             3    &\pmb{0.7876±0.0016}    & 0.8418±0.0005       & 0.8026±0.0003           & 0.7222±0.0025      &   &\pmb{0.7376±0.0011}    & 0.8229±0.0006       & 0.7323±0.0004           & 0.7061±0.0013        \\
             4    & 0.7873±0.0014        & 0.8421±0.0008       & 0.8027±0.0004           & \pmb{0.7228±0.0023}  &  & 0.7373±0.0007        & 0.8226±0.0006       & 0.7323±0.0005           & \pmb{0.7067±0.0008}             \\
             5    & 0.7874±0.0017        & 0.8419±0.0006       & 0.8026±0.0004           & 0.7223±0.0025       &  & 0.7371±0.0010        & 0.8225±0.0005       & 0.7321±0.0002           & 0.7062±0.0014        \\
    \bottomrule
    \end{tabular}
    \label{tab:impact_moe}
\end{table}

In this section, we investigate the impact of the number of experts on the performance of KT. 
The variable $e$ represents the number of experts, and we conduct experiments ranging from 1 to 5 experts. 
As shown in Table~\ref{tab:impact_moe}, we observe that as the number of experts increases, the performance of the model improves accordingly (e.g., from $e=1$ to $e=3$ in the ASSIST2009 dataset). 
Moreover, it is noteworthy that once the performance reaches its peak, the inclusion of additional experts does not significantly affect the performance.
This observation can be attributed to the fact that, when the training data size is fixed, there exists an associated optimal number of learnable experts that effectively capture the important patterns, while the additional experts may not acquire new information and therefore have minimal impact on the overall output. 
During the inference stage, in the MoE technique, the gated mechanism aggregates the outputs of all the experts to derive an overall output. 
However, despite the inclusion of additional experts, the overall output is primarily influenced by a subset of experts that exhibit strong contributions.

\subsection{Visualization of Knowledge State Estimation (RQ5)}
\label{sec:acc_interpretability}
\begin{figure*}[!hbpt]
    \centering
    \includegraphics[width=\textwidth]{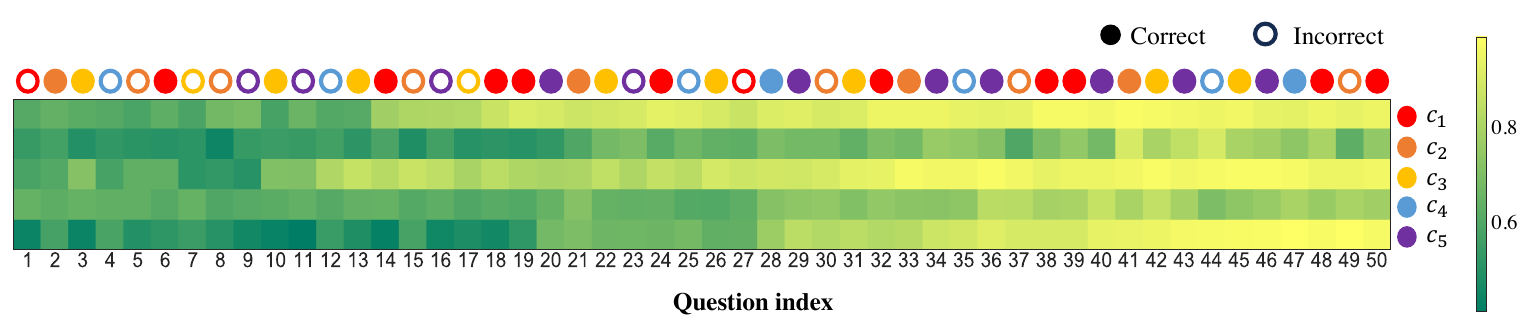}  
    \caption{The visualization of students' knowledge states in 50 questions with 5 concepts.} 
    \label{fig:predict_sequence}
\end{figure*}

To verify the accurate estimations of students' knowledge states by the Q-MCKT model, we randomly select a student sequence from Algebra2005\footnote{As its Avg. per Q is almost equal to 1.} and observe the knowledge state variations of the student in 50 questions with 5 KCs. 
From Figure \ref{fig:predict_sequence}, we observe that: (1) the estimated knowledge acquisition state of $c_2$ decreases quite a bit when the student mistakenly answers the questions that contain the KC $c_2$ twice (e.g. questions 5 and 8).
(2) On the other hand, the student gives the right answers to the questions (questions 22, 26, and 31) which are related to the KC $c_3$ hence the knowledge acquisition of KC $c_3$ is constantly increasing.
(3) As the student practices more questions, the knowledge acquisition state estimations become much more stable and after finishing all these 50 questions, the model is confident that the student has acquired the KCs $c_1$, $c_3$ and $c_5$.

\subsection{Visualization of Prediction Results (RQ6)}
\label{sec:vis_prediction_result}

\begin{figure*}[!hbpt]
    \centering
    \includegraphics[width=0.7\textwidth]{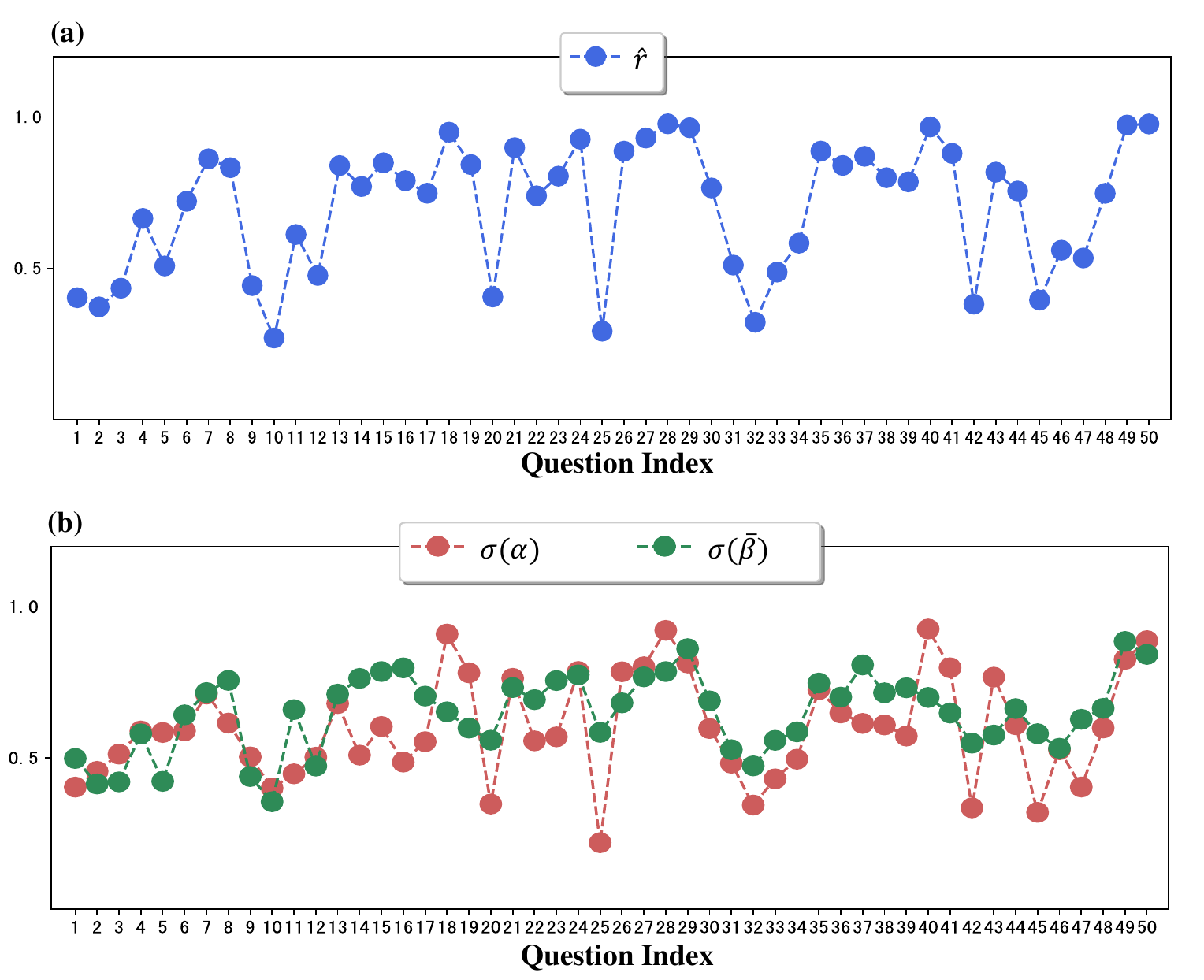}  
    \caption{The visualization of Q-MCKT's prediction results in 50 questions.} 
    \label{fig:vis_prediction}
\end{figure*}

We also randomly select a student sequence from ASSIST2009 and conduct an additional visualization to demonstrate how our Q-MCKT model produces the interpretable prediction results.
Figure~\ref{fig:vis_prediction} (a) illustrates the direct prediction result, denoted as $\hat{r}$. 
Figure~\ref{fig:vis_prediction} (b) shows the corresponding interpretable components of the prediction result $\hat{r}$, i.e., question knowledge acquisition score $\sigma(\alpha)$ and concept knowledge acquisition score $\sigma(\bar{\beta})$, generated by the Q-MCKT model.
In previous approaches, prediction results $\hat{r}$ were directly derived from the hidden states of neural networks. 
In contrast, our Q-MCKT model generates prediction results based on two interpretable components, which helps to elucidate the underlying rationale behind the prediction results.
For instance, as shown in Figure~\ref{fig:vis_prediction} (a), the model predicts a low value of $\hat{r}$ on question 25.
By examining the corresponding interpretable components in Figure~\ref{fig:vis_prediction} (b),
we observe that the student exhibits a low value of question knowledge acquisition score $\sigma(\alpha)$ and a moderate value of concept knowledge acquisition score $\sigma(\bar{\beta})$ on question 25.
This indicates that the student only has a moderate mastery of the KCs related to question 25, and the question's difficulty level is relatively high for the student acquisition.
Consequently, the model estimates that the student is likely to answer question 25 incorrectly.

\section{Conclusion}
\label{sec:conclusion}
In this paper, we propose an interpretable deep sequential KT model learning framework named Q-MCKT. 
Comparing with existing DLKT models, our Q-MCKT model is able to estimate students’ knowledge acquisition in a more robust and accurate way at both question and concept levels.
Furthermore, we design an IRT-based interpretable layer to make the Q-MCKT prediction results more interpretable. 
Through extensive experiments in four real-world datasets, we have demonstrated that Q-MCKT outperforms other state-of-the-art DLKT learning approaches, and is able to generate interpretable predictions for teachers and students.



\begin{acks}
This work was supported in part by National Key R\&D Program of China, under Grant No. 2022YFC3303600 and in part by Key Laboratory of Smart Education of Guangdong Higher Education Institutes, Jinan University (2022LSYS003).
\end{acks}

\bibliographystyle{ACM-Reference-Format}
\bibliography{tkdd2023.bib}




\end{document}